\documentclass[11pt,a4paper]{article}
\usepackage{graphicx}
\usepackage{jheppub}
\usepackage{hyperref}
\usepackage{bm}
\usepackage{amsmath}
\usepackage{amssymb}
\usepackage{amscd}
\usepackage{latexsym}
\usepackage{slashed}
\usepackage{setspace}
\usepackage{color}
\usepackage{graphicx}
\usepackage{hyperref}
\usepackage[normalem]{ulem}

\newcommand{\bs}{\begin{small}}
\newcommand{\es}{\end{small}}

\def\rs{\sqrt s}
\def\epem {e^+e^-}

\def\gappeq{\mathrel{\rlap {\raise.5ex\hbox{$>$}}
{\lower.5ex\hbox{$\sim$}}}}

\def\lappeq{\mathrel{\rlap{\raise.5ex\hbox{$<$}}
{\lower.5ex\hbox{$\sim$}}}}

\title{Top pair production at a future $e^+e^-$ machine \\
in a composite Higgs scenario }

\date{\today}
\author[a]{D. Barducci,}
\author[b]{S. De Curtis,}
\author[c,d]{S. Moretti}
\author[e]{and G.M. Pruna}

\affiliation[a]{LAPTh, Universit\'e Savoie Mont Blanc, CNRS, 9 Chemin de Bellevue, B.P. 110, F-74941 Annecy~le-Vieux,~France}
\affiliation[b]{INFN, Sezione di Firenze, and Dept. of Physics and Astronomy, University of Florence, Via G.~Sansone 1, 50019 Sesto Fiorentino, Italy}
\affiliation[c]{School of Physics and Astronomy, University of Southampton, Highfield, Southampton SO17~1BJ,~U.K}
\affiliation[d]{Particle Physics Department, Rutherford Appleton Laboratory, Chilton, Didcot, Oxon~OX11~0QX,~UK}
\affiliation[e]{Paul Scherrer Institut,
CH-5232 Villigen PSI, Switzerland}

\abstract{The top quark plays a central role in many New Physics scenarios and in understanding the details of
Electro-Weak Symmetry Breaking. In the short- and mid-term future, top-quark studies will mainly be driven by the experiments at the Large Hadron Collider. Exploration of top quarks will, however, be an integral part of particle physics studies at any future facility and an $e^+ e^-$ collider will have a very comprehensive top-quark physics program. 
We discuss the possibilities of testing NP in the top-quark sector 
within a composite Higgs scenario through deviations from the Standard Model in top pair production for different Centre-of-Mass energy options of a future $e^+e^-$ machine. In particular, we focus on precision studies of the top-quark sector at a CM energy ranging from 370 GeV up to 3 TeV.}

\emailAdd{barducci@lapth.cnrs.fr}
\emailAdd{decurtis@fi.infn.it}
\emailAdd{s.moretti@soton.ac.uk}
\emailAdd{giovanni-marco.pruna@psi.ch}

\begin{document}
 \begin{flushright}
   LAPTH-020/15    \\
   PSI-PR-15-05    \\
 \end{flushright}
\maketitle

\section{Introduction}
The large mass gap between the third generation Standard Model (SM) quarks and the first two, possibly hints at an intrinsic difference between the nature of these particles. While there is
 no explanation for this mass hierarchy in the SM, several New Physics (NP) scenarios attempt to find a resolution to this
puzzle. Among these, Composite Higgs Models (CHMs), wherein the Higgs boson emerges as a pseudo Nambu-Goldstone Boson (pNGB) of a spontaneous breaking of a global symmetry in a new strongly interacting sector, play a special role.
In fact, this type of NP framework is
primarily designed as an alternative to the Electro-Weak Symmetry Breaking (EWSB) pattern of the SM
and the idea goes back to the '80s~\cite{Kaplan:1983fs,Georgi:1984ef,Georgi:1984af,Dugan:1984hq}. However, one modern ingredient 
of CHMs is the mechanism of so-called `partial compositeness'~\cite{Kaplan:1991dc}, wherein the heaviest SM fermions
(indeed, the third generation states) 
mix with new states arising from the strong sector in order to explain the mass difference with respect to the other SM fermions.

The simplest example, based on the symmetry breaking pattern $SO(5)\to SO(4)$, was considered in~\cite{Agashe:2004rs} in the context of 5-Dimensional (5D) scenarios while deconstructed 4D effective descriptions of this model were recently proposed~\cite{Panico:2011pw,DeCurtis:2011yx}.
These explicit CHM constructions present several features of phenomenological relevance at colliders, as they include in their spectrum only the lowest lying resonances, both spin 1/2 and spin 1 states, arising as bound states of the new strongly interacting dynamics, that are the only degrees of freedom which might be accessible at the Large Hadron Collider (LHC) and future $e^+e^-$ machines.
In particular the 4D Composite Higgs Model (4DCHM) of~\cite{DeCurtis:2011yx} allows, due to the choice of its fermionic sector, for a finite one loop Higgs potential, calculable via the Coleman-Weinberg technique~\cite{Coleman:1973jx}, and, for a natural choice of the model parameters, the emerging Higgs mass turns out to be compatible with the measurements by the ATLAS~\cite{Aad:2012tfa} and CMS~\cite{Chatrchyan:2012ufa} experiments.

Part of the new particle content of CHMs, both fermions and vector bosons, can manifest itself in $t\bar t$ production processes in two ways.
One the one hand, potentially large deviations from the SM value of the $Ztt$ coupling can arise, because of the mixing between the top quark and its partners ($t'$s) or else between the $Z$ and its partners ($Z'$s). On the other hand, the neutral $Z'$ states
 can enter as propagating particles in the diagrams describing top pair production and thus contribute on their own or else through interference effects with the SM neutral gauge bosons. 

Concrete CHM realisations
can therefore be used as an ideal theoretical scenario where to test the composite Higgs idea in $t\bar t$ production.
This is the reason why a close look at this production process within the 4DCHM was taken in Ref.~\cite{Barducci:2012sk}. Herein, it was
shown that the additional $Z'$ resonances present in this CHM scenario could greatly affect the phenomenology of $t\bar t$ production at the LHC, 
through their effects onto both the cross section and (charge and spin) asymmetries, so long that they can be produced
resonantly and subsequently decay into top-antitop pairs. Indeed, Ref.~\cite{Barducci:2012sk}
confirmed that the 13 TeV stage of the LHC will enable one to even detect such $Z'$ states, assuming standard detector performance and machine luminosity. 

However, at the LHC, the subprocess induced by $Zt\bar t$ interactions in top-pair production
 is EW in nature and further induced by $q\bar q$ scattering, hence it sees a double suppression with respect to the
leading QCD background due to the $gg\to t\bar t$ channel.
It cannot therefore be used to investigate deviations in the $Zt t$ coupling for which one should rely 
instead on the $t\bar t Z$ final state, whose experimental precision is unfortunately extremely limited at 
present, with errors of order 100\% after the 7, 8 TeV LHC runs with full luminosity~\cite{ttZ,ttZ-ATLAS,ttZ-ICHEP2014,Khachatryan:2014ewa}, and unlikely to be improved significantly at Run 2 with standard machine configuration\footnote{The sensitivity to variations of the $Zt t$ coupling through one-loop EW effects entering both $gg$ and $q\bar q\to t\bar t$ in the full 4DCHM was found to be negligible by a 4DCHM remake of
the calculation of Refs.~\cite{Moretti:2006nf,Moretti:2012mq}.}. Possibly, a tenfold LHC luminosity upgrade~\cite{Gianotti:2002xx} could overcome the poor sensitivity to the $Zt t$ coupling of the 
current CERN machine. Needless to say,
a future $e^+e^-$ collider operating at the
$t\bar t$ threshold and above would therefore be the ideal environment where to test both real and virtual CHM effects
in the top-antitop cross section and asymmetries. 

Such machine prototypes
currently include the International Linear Collider (ILC) 
\cite{Phinney:2007gp,Behnke:2007gj,Phinney:2007zz,Djouadi:2007ik,Brau:2007zza,BrauJames:2007aa,Behnke:2013lya,Adolphsen:2013kya,Adolphsen:2013jya,Behnke:2013xla,Baer:2013cma},
the Compact Linear Collider (CLIC)~\cite{Aicheler:2012bya} and the Triple Large Electron-Positron (TLEP) collider~\cite{Gomez-Ceballos:2013zzn} (whose design study recently merged in the Future Circular Collider (FCC) one with the electron-positron option (FCC-ee)~\cite{FCC-ee}). Rather independently of the actual machine, there are common characteristics that render $\epem$ colliders attractive with 
respect to $pp$ and $p\bar p$ ones:
(i) a well defined initial state and controllable collision energy $\rs$;
(ii) a particularly clean experimental environment since all processes are initiated by EW interactions;
(iii) more accurate theoretical predictions for all event rates for the same reason;
(iv) signals that are in general not swamped by backgrounds as the rates for the former do not differ from those of the latter by orders of magnitudes, as it happens at hadronic machines because of the QCD interaction dominance;
(v) the possibility (at least in principle) of reconstructing any hadronic final state, with accuracy mainly driven by the detector jet-energy resolution and flavor-tagging efficiency;
(vi) the possibility of having highly-polarised initial beams, which give an extra handle for enhancing particular signals and suppressing their backgrounds too.
As a consequence, even at moderate Centre-of-Mass (CM) energy $\rs$, an $\epem$ collider can be quite competitive with respect to hadron colliders in performing precision physics, in particular, in the Higgs-boson sector, providing model-independent measurements of couplings and covering with ease even hadronic decay channels, and in the top-quark system, enabling the precise extraction of the top mass, width and couplings. 

The CM energy is of course a critical parameter in the present discussion. ILC studies have been focused on the $\rs=500$ GeV option for quite a long time, 
while envisaging the possibility to run at the top-pair threshold $\rs\simeq 350$ GeV for top precision measurements as well as even smaller $\rs$ in order to scrutinise $W$ and $Z$ boson physics. 
However, after the Higgs boson discovery, it has become obvious that a CM energy that enhances the associated production $\epem\to HZ$ would be optimal for Higgs precision measurements, which are now the highest priority. Both ILC and TLEP (and possibly CLIC)
would conform with this physics requirements. Nevertheless, going to higher $\rs$, above and beyond 500 GeV, remains in the plan, as one can enlarge considerably the physics scope of the machine, by not only complementing 
the potential of studying the Higgs boson sector by onsetting Higgs production via vector boson fusion, but also enabling studies of new objects that may be found at the LHC in the meantime. While the possibility for upgrades to higher energies (up to $\rs\sim1$ TeV) with the same ILC technology is also envisaged, the multi-TeV regime is the natural operating ground of CLIC. 

It is the purpose of this study to assess the sensitivity of an $e^+e^-$ machine operating at energy scales between $\sim 2m_t$ and 3 TeV to variations of $t\bar t$ production induced by a composite Higgs scenario, by looking at inclusive observables, such as cross section as well as (charge and spin) asymmetries, with and without beam polarisation, and differential ones, mapped in terms of energy and angular variables.
The plan of the paper is as follows. In Sect.~\ref{top-measure} we describe the top-quark measurements foreseen at a future $e^+e^-$ machine and contrast these with those planned at the LHC.
In the following part, Sect.~\ref{top-4DCHM}, we detail the effects onto the $t\bar t$ system induced by the 4DCHM while in Sect.~\ref{top-selection} we describe how they can be extracted. 
We then summarise and conclude in Sect.~\ref{conclude}.

\section{\label{top-measure} Top-quark measurements}

The top quark is the heaviest known elementary particle.
Thanks to its large mass, and the related strength of its coupling to the Higgs boson, this particle plays a central role in many NP models, particularly in understanding the details of EWSB. The properties of the top quark have in fact profound consequences, for instance, on the stability of the SM vacuum, which depends on the precise value of the top mass.
As mentioned, while in the short- and mid-term future top-quark studies will be mainly driven by the LHC experiments, they will be an integral part of particle physics studies at any future facility~\cite{Agashe:2013hma,Moortgat-Pick:2013awa}.
An $e^+ e^ -$ collider will have a rich top-quark physics programme mainly in two domains: accurate determination of top properties at the $t \bar t$ production threshold (e.g., measurement of $m_t$ with an uncertainty of $\sim$100 MeV on the ${\overline{\rm MS}}$ mass and of the top-quark gauge couplings at the percent level) and search for NP with top quarks above the $t\bar t$ threshold (e.g., direct measurement of the top-quark Yukawa coupling, extraction of rare top-quark decays, searches for new particles decaying into top-quark pairs, etc.). 

\begin{figure}[!t]
\begin{center}
\hspace{-0.5cm}
\includegraphics[width=0.48\textwidth]{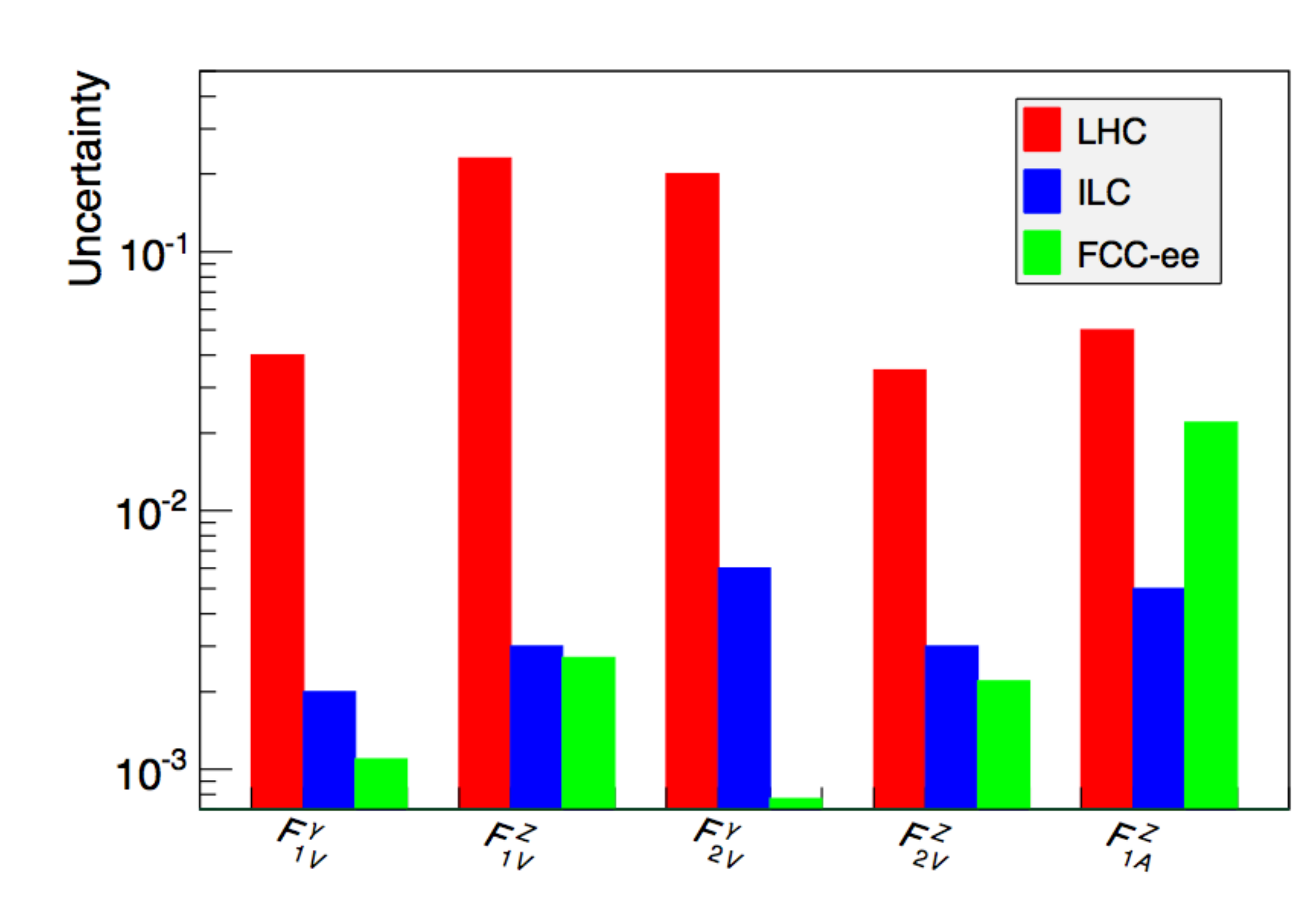}
\includegraphics[width=0.48\textwidth]{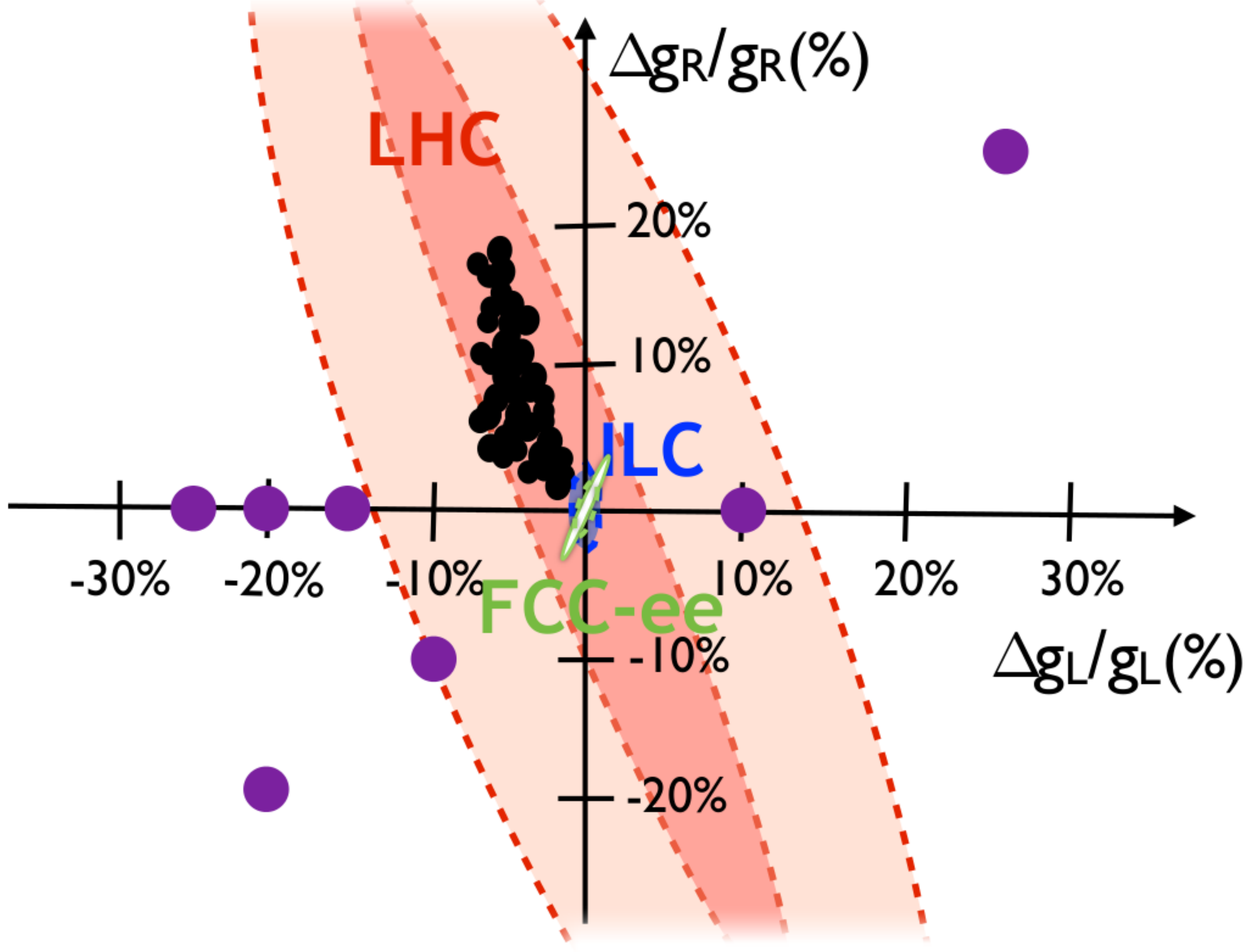}
\vspace{-0.1cm}

\caption{\small \it Left: Statistical uncertainties on the CP-conserving top-quark axial and vector form factors expected at the LHC-13 with 300 fb$^{-1}$~\cite{Juste:2006sv,Rontsch:2014cca} (red), at ILC-500 with 500 fb$^{-1}$ and beam polarisation of $P=\pm 0.8$, $P'=\pm 0.3$ (from~\cite{Baer:2013cma},~\cite{Amjad:2013tlv,Asner:2013hla}) (blue) and at TLEP (FCC-ee) without beam polarisation with $\sqrt s=360$ GeV and 2.6 ab$^{-1}$~\cite{Janot:2015yza} (green). Right: Typical deviations for the $Zt_L t_L$ and $Z t_Rt_R$ couplings in various NP models represented by purple points (see~\cite{Richard:2014upa}). The black points indicate the expected deviations for different choices of the 4DCHM parameters (see Fig.~\ref{tLR} later on). Also shown are the sensitivities expected after LHC-13 with 300 fb$^{-1}$, (region inside the red-dashed lines), after HL-LHC with 3000 fb$^{-1}$ (region inside the inner red-dashed lines), from ILC-500 with polarised beams (region inside the blue-dashed lines)~\cite{Baer:2013cma} {and from FCC-ee (region inside the green lines: the continuous(dashed) line indicates the bounds extracted from the angular and energy distribution of leptons($b$-quarks))} in the extraction of the $Zt_L t_L$ and $Z t_Rt_R$ couplings.}
\label{grojean}
\end{center}
\end{figure}

Moreover, machines with the possibility of polarised electron and/or positron beams will allow additional precision measurements of the couplings of the top quark to the photon and to the $Z$ by measuring polarisation-dependent asymmetries and cross sections~\cite{Grzadkowski:2000nx}. The ILC (and possibly CLIC), for example, provides an ideal setup for this type of scenario, with a large set of observables allowed by the polarisation of both leptonic beams. In fact, also a machine like TLEP (or a generic FCC-ee) will have the possibility to extract with great precision the couplings of the photon and $Z$ to the top quark even without polarised initial beams, by analysing observables involving different helicities of the produced (anti)top
quark~\cite{Janot:2015yza}, as its polarisation is transferred to the final state particles via the weak decays rather efficiently. 
In the end, the two beam options (with and without polarisation) are competitive. On the one hand, the polarisation 
of the initial $e^+e^-$ state does not correspond to a unique spin combination of the final $t\bar t$ system. On the other hand, the lack of beam polarisation is compensated by a larger integrated luminosity.

The top-quark couplings to the photon and $Z$ can be parametrised in several ways. For example, in~\cite{Baer:2013cma}, the analysis makes use of the usual form factors defined by\footnote{The most general Lorentz-invariant vertex function describing the interaction of a neutral vector boson $X$ with two top quarks can be written in terms of ten form factors which reduce to four when both top quarks are on-shell (hence, this is true for $t\bar t$ but not for $t\bar t X$ production).}:
\begin{equation}
\Gamma_\mu^{ttX}(k^2,q,\bar q)=-i e\big [\gamma_\mu (F_{1V}^X(k^2)+\gamma_5F_{1A}^X(k^2))+\frac {\sigma_{\mu\nu}}{2 m_t}(q+\bar q)^{\nu}(i F_{2V}^X(k^2)+\gamma_5F_{2A}^X(k^2))\Big]
\label{F}
\end{equation}
where $e$ is the proton charge, $m_t$ is the top-quark mass, $q$ ($\bar q$) is the outgoing top (antitop) quark four-momentum and $k^ 2=(q+\bar q)^2$. The terms $F_{1V,A}^{X}(0)$ in the low energy limit are the $ttX$ vector and 
axial-vector form-factors (at tree level the only non-vanishing ones in the SM are $F_{1V,A}^{Z}$ and $F_{1V}^\gamma$). 

It is particularly interesting to study the couplings of the top quark to the photon and $Z$ to search for NP effects. As
intimated, at hadron colliders, the EW production of $t\bar t$ is suppressed with respect to QCD production and this is especially true at the LHC where most of the $t\bar t$ production comes from gluon-gluon fusion. In the case of 
the $t\bar t Z$ final state, relatively clean measurements are expected at the LHC for the $Z$ decaying leptonically. However the cross section is quite small, so that only a precision of about 10\% for $F_{1A}^Z$ is expected after a few 100 fb$^{-1}$ (at the HL-LHC with an integrated luminosity of 3000 fb$^{-1}$ the precision of this measurement is expected to improve by a factor 3 or so). In contrast, the sensitivity to $F_{1V}^Z$ is very poor. The sensitivities achievable for the CP-conserving top-quark form factors $F^X_{1V},~F^Z_{1A},~ F^X_{2V}$, defined in Eq.~(\ref{F}), at the LHC running at 13 TeV 
(LHC-13) after 300 fb$^{-1}$ are reported in the left panel of Fig.~\ref{grojean} and are taken from~\cite{Juste:2006sv,Rontsch:2014cca}. Also shown are the sensitivities reachable at the 500 GeV ILC (ILC-500) assuming an integrated luminosity of 500 fb$^{-1}$ and 80\% electron and 30\% positron polarisation~\cite{Amjad:2013tlv,Asner:2013hla}. Very recently an analysis in~\cite{Janot:2015yza} exploited the reach of TLEP (FCC-ee) in the energy configuration $\sqrt{s}$=360 GeV with unpolarised beams and with an integrated luminosity of 2.6 ab$^{-1}$. The result is that a choice of optimal observables of the lepton angular and energy distributions of events from $t\bar t$ production in the $l \nu q \bar q b \bar b$ final states can give sensitivities to the top-quark EW couplings which are comparable to the ones from a polarised ILC-500. {This is well manifest from the two green contours in the figure, wherein the plain ellipse is obtained via the angular and energy distributions of leptons while the dashed one is extracted from the angular and energy distributions of $b$-quarks.} Overall, {despite relatively merits of ILC versus TLEP (FCC-ee) cannot (and need not) be rigorously ascertained here\footnote{{For example, in our qualitative exercise we have paid little or no attention to the use of appropriate correlation matrices in the two cases.}},} it is clear that an $e^+ e^-$ collider is very powerful in increasing the sensitivities to both the vector and axial-vector form factors beyond the LHC scope, thus allowing for an independent measurement of the left- and right-components of the $Z t t$ couplings at the percent level.

Many extensions of the SM typically induce large deviations in $Z$ boson couplings to $t \bar t$. In general, any new fermion which mixes with the third generation quarks might affect this coupling strength and, depending on the NP scheme, such deviations can be different for the left- and right-components. In~\cite{Richard:2014upa} a wide spectrum of predictions for deviations of the $Ztt$ couplings in various BSM scenarios, like Randall-Sundrum models, Little Higgs and CHMs, was considered. In all these schemes the top quark is assumed to carry a great deal of compositeness through the mixing with new heavy fermions but also a mixing between the SM gauge bosons and the heavy vector states can induce variations of the top-quark EW couplings. These deviations, expressed in the $ Z t_L t_L$ and $Zt_R t_R$ couplings, for a generic NP scale around 1 TeV, are pictorially drawn in Fig.~\ref{grojean} (right panel), where the purple points represent typical deviations for different NP scenarios. The region within the {outer(inner) red-}dashed lines represents the sensitivity which can be reached by the LHC-13{(HL-LHC)} with 300{(3000)} fb$^{-1}$ of integrated luminosity through the $ttZ$ cross section measurement~\cite{Juste:2006sv,Rontsch:2014cca}, while the ILC-500 sensitivity with 500 fb$^{-1}$ of integrated luminosity (detailed by the blue-dashed contour) can be at the percent level (the ILC-500 expected accuracies for the $t_L$ and $t_R$ couplings can be estimated to be 
$\Delta (Zt_Lt_L)/Zt_Lt_L(\%)\simeq0.6$ and $\Delta (Zt_Rt_R)/Zt_Rt_R(\%)\simeq1.4$~\cite{Amjad:2013tlv,Asner:2013hla}). 
As a result, with an $e^+e^-$ collider, one can in general reach an excellent separation of different NP models while the LHC will not be able to do so even at the high-luminosity option. This is clear for the NP scheme we will consider in this paper as a prototype of CHMs. In fact, in the right plot of Fig.~\ref{grojean} we also include the points corresponding to the deviations obtained within the 4DCHM (as described in the next section). It is then clear that the LHC will not be able to disentangle the 4DCHM by the SM for a wide range of its parameter space\footnote{Moreover, at any $e^+ e^-$ collider, all top EW couplings, including $Wtb$, will be measured allowing for a full separation between axial and vector couplings and between the $Ztt$ and $\gamma tt$ components.}.

\section{\label{top-4DCHM} Top pair production in the 4DCHM}

\subsection{Calculation}

Let us consider the process $e^+e^-\to t \bar t$ in the calculable setup provided by the 4DCHM~\cite{DeCurtis:2011yx}, which encodes the main characteristics of CHMs wherein the physical Higgs state is a pNGB 
emerging from the $SO(5)\to SO(4)$ breaking. Models of this kind generally predict modifications of its coupling to both bosons and fermions of the SM, hence the measurements of these quantities represent a powerful way to test the possible non-fundamental nature of the newly discovered state~\cite{Barducci:2013wjc,Barducci:2013ioa}. Furthermore, the presence of additional particles, both spin-1 and spin-1/2, in the spectrum of such CHMs leads to mixing effects with the 
SM states with identical quantum numbers as well as new Feynman diagram topologies, both of which would represent a source of deviations from the SM expectations. 

Now, focusing onto the 4DCHM as an illustrative example representing a description of the minimal additional heavy matter to ensure a finite effective potential for the Higgs state, modifications in the $e^+e^-\to t \bar t$ process arise via the following effects: (i) corrections to the $Zee$ coupling due to the mixing of the $Z$ with the $Z'$s, which however is very small in the parameter range considered; (ii) corrections to the $Z t t$ coupling which come from the mixing of the $Z$ with the $Z'$s but also from the mixing of the top (antitop) with the extra-(anti)fermions, as expected because of the partial compositeness mechanism; (iii) the presence of new particles,
namely the $s$-channel exchange of $Z'$s, which can be sizeable also for large $Z'$ masses due to the interference with the SM states. In the model independent effective approaches which are generally used for the phenomenological study of CHMs, the latter effect is not captured. However, as we will show, it can be crucial from moderate to high CM energies of a lepton collider. 
Specifically, the framework on which we base our analysis, the 4DCHM, describes, in addition to the SM particles, a large number of new ones both in the fermionic and bosonic (gauge) sector.
In particular, 5 neutral and 3 charged extra spin-1 resonances are present, together with 8 partners of the top and of the bottom quarks, called in a general way $t^\prime$ and $b^\prime$. Moreover 4 exotic extra fermions, 2 with charge 5/3 and 2 with charge $-4/3$, are present (called $X$ and $Y$, respectively). See~\cite{Barducci:2012kk} for details of the
model implementation adopted here.

In general, the 4DCHM can be schematised as a two site model arising from an extreme deconstruction of a 5D theory and can be described by two sectors, mixed between themselves via the mechanism of partial compositeness. 
The gauge structure of the elementary sector is associated with the $SU(2)_L\otimes U(1)_Y$ SM gauge symmetry whereas the composite sector has a local $SO(5)\otimes U(1)_X$ symmetry. The parameters of the gauge sector are: $f$, the scale of the spontaneous global symmetry breaking $SO(5)\to SO(4)$ in the TeV range, and $g_\rho$, the $SO(5)$ strong gauge coupling constant (which for simplicity we take equal to the $U(1)_X$ one).
Regarding the fermionic sector, we just recall that the new heavy states (20 in total) are embedded in fundamental representations of $SO(5)\times U(1)_X$ and two multiplets of states for each of the SM third generation quarks are introduced in such a way that only top and bottom quarks mix with these heavy fermionic resonances in the spirit of partial compositeness. This choice of representation is a realistic scenario compatible with EW precision measurements. The SM third generation quarks are embedded in an incomplete representation of $SO(5)\otimes U(1)_X$ in such a way that their correct hypercharge is reproduced via the relation $Y=T^{3R}+X$. Notice also that this is the minimum amount of new heavy fermion content necessary for an ultra-violet finite Higgs potential which is radiatively generated~\cite{DeCurtis:2011yx}. The Lagrangian describing the gauge and top sectors of the 4DCHM is the following:
\begin{equation}
\begin{split}
\mathcal{L_{\rm 4DCHM}}\supset&\frac{f^2}{2}Tr|D_{\mu}\Omega|^2+f^2(D_{\mu}\Phi)(D_{\mu}\Phi)^T+\\
+&(\Delta_{t_L}\bar{q}_L\Omega\Psi_T+\Delta_{t_R}\bar{t}_R\Omega\Psi_{\tilde{T}}+h.c.)+\\
-&m_*(\bar{\Psi}_T \Psi_T+\bar{\Psi}_{\tilde{T}}\Psi_{\tilde{T}})+\\
-&(Y_T\bar{\Psi}_{T,L}\Phi^T\Phi\Psi_{\tilde{T},R}+M_{Y_T}\bar{\Psi}_{T,L}\Psi_{\tilde{T},R}+h.c.)\\
\end{split}
\end{equation}
with the covariant derivatives defined by
\begin{eqnarray}
D^{\mu}\Omega&=&\partial^{\mu}\Omega-i g_{0}\tilde{W}\Omega+i g_\rho\Omega\tilde{A},\\
D_{\mu}\Phi&=&\partial_{\mu}\Phi-i g_\rho\tilde{A}\Phi
\end{eqnarray}
where $(g_0,\tilde W)$ and $(g_\rho,\tilde A)$ indicate in a generalised way the gauge couplings and gauge fields of the elementary and composite sector, respectively.
The field $\Omega$, responsible for the symmetry breaking, is given by
\begin{equation}
\Omega=\textbf{1}+i\frac{s_h}{h}\Pi+\frac{c_h-1}{h^2}\Pi^2,\quad s_h,c_h=\sin,\cos(h/2 f),\quad   h=\sqrt{h^{\hat{a}}h^{\hat{a}}},
\end{equation}
where $\Pi=\sqrt{2}h^{\hat{a}}T^{\hat{a}}$  is the PNGB matrix with $T^{\hat{a}}$ the broken generators of $SO(5)/SO(4)$. The field $\Phi$ is a vector of $SO(5)$  that describes the spontaneous symmetry breaking of $SO(5)\rightarrow SO(4)$ and is given by $\Phi=\phi_0\Omega^T$ where $\phi_0^i=\delta^{i5}$.
Finally, $\Psi_{T,\tilde T}$ are the two fundamental representations of $SO(5)$ in which the new fermions are embedded in, and $\Delta_{t_L,t_R},m_*$ and $Y_T,M_{Y_T}$ the parameters describing the linear mixing between the elementary and composite sector, the mass parameter of the new extra quarks and the parameters describing the new quark interactions with the GB fields and among themselves. For brevity we do not write the Lagrangian of the bottom sector, for which similar expressions hold.

Among the various new states of the 4DCHM, the relevant ones, for the purpose of our study, are the neutral gauge bosons $Z'_{2,3}$ (in fact the $Z'_{1,4}$ states are essentially inert as they do not couple to $e^+e^-$ while the $Z'_5$ is only poorly coupled~\cite{Barducci:2012kk}) and the additional fermions which affect the $Z'$ widths. The mass spectrum of the spin-1 fields is computed at the scale $f$, with $f$ around 1 TeV to avoid excessive fine tuning. In particular, $Z'_{2,3}$ are nearly degenerate, while the $Z'_5$ is heavier:
\begin{equation}
 \begin{split}
& M^2_{Z^\prime_2}\simeq \frac{ m_\rho^2}{c_\psi^2} (1-\frac{s_\psi^2 c_\psi^4}{4 c_{2\psi}}\xi),\\
& M^2_{Z^\prime_3}\simeq \frac{ m_\rho^2}{c_\theta^2} (1-\frac{s_\theta^2 c_\theta^4}{4 c_{2\theta}}\xi),\\
& M^2_{Z^\prime_5}\simeq 2 m_\rho^2 \left[1+\frac 1 {16} (\frac 1 {c_{2\theta}}+\frac 1{2 c_{2 \psi}})\xi\right]
\end{split}
\label{MZ'}
\end{equation}
with $m_\rho=f g_\rho$, $s_\theta/c_\theta=g_0/g_\rho$, $s_\psi/c_\psi=g_{0Y}/g_\rho$, $g_0$ and $g_{0Y}$ 
being the $SU(2)_L$ and $U(1)_Y$ gauge couplings respectively, $\xi=v^2/f^2$ and $v$ the Vacuum Expectation Value (VEV) of the Higgs state. The first two resonances correspond the the neutral component of the $(\textbf{3,1})$ and $(\textbf{1,3})$ triplet of the unbroken $SU(2)_L$ and $SU(2)_R$ while the latter to the heaviest gauge boson of the $SO(5)/SO(4)$ coset.

\begin{figure}[!t]
\begin{center}
\hspace{-0.5cm}
\includegraphics[width=0.35\textwidth]{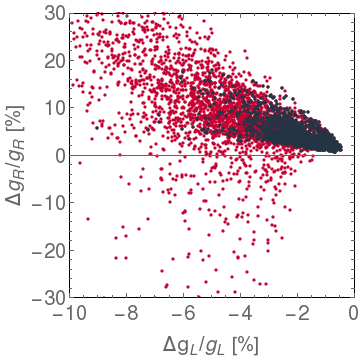}\hspace{1.1cm}
\includegraphics[width=0.35\textwidth]{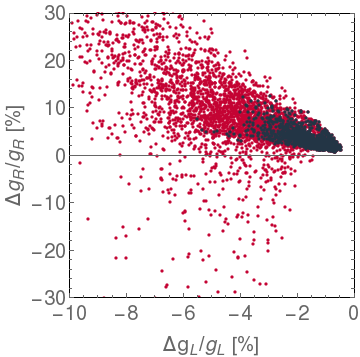}\\
\vspace{-0.1cm}
\caption{\small \it 
Deviations of the left- and right-handed couplings of the $Z$ to $t \bar t$ in the 4DCHM with respect to the SM ones. The red points correspond to a scan with $0.75$ TeV $\le f\le 1.5$ TeV, $1.5\le g_\rho\le 3$ and on the extra-fermion sector parameters as described in~\cite{Barducci:2012kk}. The black points correspond to the subset with $M_{Z^\prime}\sim f g_\rho>2$ TeV
and $M_{t^\prime}>782$ GeV, $M_{b^\prime}>785$ GeV and $M_{X}>800$ GeV (left), while in the right plot bounds on the masses of the extra fermions are imposed to be universal and equal to 1 TeV. }
\label{tLR}
\end{center}
\end{figure}

The fermionic Lagrangian of the 4DCHM~\cite{DeCurtis:2011yx} contains the mixing parameters relating the elementary and the composite sectors as well as the Yukawas of the latter. The top and bottom quark masses are proportional to the EWSB parameters and to the elementary/composite sector mixings as suggested by the partial compositeness hypothesis. 
For the process of interest here, $e^ + e^ - \to\ \gamma, Z, Z'\to t\bar t$, we need the couplings of the $Z,Z'$ to the electron-positron pair which live in the elementary sector and those to the top-antitop pair which interacts with the composite fermionic sector. While the former come only from the mixing $Z-Z'$, in the latter also the mixing of the third generation (anti)quarks with the new heavy fermions has to be taken into account.
The analytic expressions at the leading order in $\xi$ for the neutral current interaction Lagrangian of the 4DCHM are given in~\cite{Barducci:2012sk}. Concerning the $Ztt$ vertex modifications, we plot in Fig.~\ref{tLR} the deviations of the left- and right-handed couplings with respect to the SM ones. (Here, $\Delta g_L/g_L=(g_L^{\rm 4DCHM}-g_L^{\rm SM})/g_L^{\rm SM}$, where $g_L^{\rm SM}$ is the $Z t_L t_L$ coupling within the SM, and the same definition holds for the right-component.)
%
In the plots, the red points correspond to $f=0.75-1.5$ TeV, $g_\rho=1.5-3$ (which are natural values for the scale $f$ of the underlying strong sector, in order to avoid the aforementioned fine tuning, and for the new strong coupling constant) and a scan on the extra-fermion sector parameters as described in~\cite{Barducci:2012kk}. In doing so, we impose
 the requirement that the physical quantities $e, M_Z, G_F, m_t,m_b,v,m_H$ are consistent with experimental data.
For the first three we adopt the current Particle Data Group (PDG) values~\cite{Beringer:1900zz} while for the last four we require
$165\le m_t\le 175$, $2$ GeV $\le m_b \le 6$ GeV, $v \simeq 246$ GeV and $120$ GeV $\le m_H\le 130$ GeV.
The black points (already pictorially reported in Fig.~\ref{grojean}) 
 correspond instead to $M_{Z^\prime}\sim f g_\rho>2$ TeV
and $M_{t^\prime}>782$ GeV, $M_{b^\prime}>785$ GeV and $M_{X}>800$ GeV for the left panel. These requests represent allowed configurations from both EWPTs\footnote{While a complete calculation of the EW oblique observables is beyond the scope of this work, these choices of $f$ and $m_{X,T,B}$ have been made following the guidance of \cite{Grojean:2013qca}, and can be considered as a safe estimation, in order to satisfy EWPTs.} and direct searches (see~\cite{Chatrchyan:2013uxa,CMS:2013una,Chatrchyan:2013wfa} for the latest CMS results for direct searches of extra fermions).
For the case of the $t^\prime$ and $b^\prime$ states, the experimental collaborations set bounds depending on their decay branching fractions. For our analysis we have chosen to impose the strongest of these limits as a naive estimate.

The 13 TeV run of the LHC is expected, if no such $t'$ and $b'$ states are discovered, to set stronger bounds on the masses of these particles. For this reason we enforce in the right panel of Fig.~\ref{tLR} a general 1 TeV cut on the masses of all the extra fermions. This requirement slightly reduces the number of allowed points without however significantly changing the size of the deviations. Therefore our analyses will be performed while enforcing the actual bounds on the extra fermions.

As it is clear from the plots, there is a common trend in reduction of the left-handed coupling while the right-handed one is enhanced. The modifications can be substantial: up to 20\% for the $Zt_R t_R$ coupling whereas they can reach $-10\%$ for the $Zt_L t_L$ one.
By comparing with the sensitivities shown, one can realise that the 4DCHM induced deviations may be hard to detect 
at the LHC-13 or HL-LHC. In fact, the LHC can reach a 10\% sensitivity in the $Ztt$ axial coupling measurement while it is quite insensitive to the vector one~\cite{Baer:2013cma}. On the contrary, the ILC expected accuracies~\cite{Amjad:2013tlv,Asner:2013hla} could be able to disentangle the 4DCHM (or similar CHMs) from other NP models and the SM itself.

As stated, the indirect signals from a CHM are not only encoded in the coupling modifications. In fact, the interferences due to the $Z'$s (mainly $Z'_{2,3}$) $s$-channel exchanges play a crucial role. In this respect it is important to stress that the extra gauge bosons of the 4DCHM may have different widths depending on the choice of the 
composite fermion sector parameters. As pointed out in~\cite{Barducci:2012kk}, one can divide the mass spectrum of the latter
in two distinct configurations, as follows: (i) a regime where the mass of the lightest fermionic resonance is too heavy to allow for the decay of a $Z'$ in a pair of heavy fermions and, consequently, the widths of the $Z'$s are small, typically well below 100 GeV; (ii) a regime where a certain number of masses of the new fermionic resonances are light enough to allow for the decay of a $Z'$ in a pair of heavy fermions and, consequently, the widths of the involved $Z'$ states are relatively large and can become even comparable with the masses themselves.

Let us now compare the deviations due to the 4DCHM with respect to the SM expectations for the following observables: the total cross section $\sigma(e^+e^-\to t\bar t)$, the Forward-Backward Asymmetry $A_{FB}$ plus the (single and double, respectively) spin asymmetries $A_{L}$ and $A_{LL}$ defined as follows:
\begin{equation}
\begin{split} 
& A_{FB}= \frac{N(\cos\theta^*>0)-N(\cos\theta^*<0)}{N_{\rm tot}},\\
& A_L=\frac{N(-,-)+N(-,+)-N(+,+)-N(+,-)}{N_{\rm tot}},\\
& A_{LL}=\frac{N(+,+)+N(-,-)-N(+,-)-N(-,+)}{N_{\rm tot}},
\end{split}
\label{asym}
\end{equation}
with $\theta^*$ the polar angle in the $t \bar t$ rest frame (which would coincide with the CM frame if no
radiation from the initial state occurred prior to the hard scattering). $N$ denotes the number of observed events 
in a given hemisphere (for $A_{FB}$) or for a set polarisation (for $A_L$ and $A_{LL}$), in which case
 its first (second) argument corresponds to the helicity of the final state top (antitop), whereas $N_{\rm tot}$ is the total number of events. 
In particular,
$A_L$ singles out one final state particle, comparing the number of its positive and negative helicities while summing over the helicities of the antiparticle (or \emph{vice versa}) whereas $A_{LL}$ relies on the helicity flipping of either of the final state particles. These spin (or polarisation) asymmetries focus on the helicity structure of the final state fermions which can be reconstructed in top (antitop) (semi-)leptonic decays with leptons used as spin analysers. In general, such asymmetries are extracted as coefficients in the angular distributions of the top (antitop) decay products, as described in~\cite{Bernreuther:2008ju}.

In order to perform the aforementioned calculations in an automated way, several openly available tools were used.
The described model was
first implemented in LanHEP~v3.1.9~\cite{Semenov:2010qt} with the use of the SLHA+ library~\cite{Belanger:2010st},
so as to obtain an output of Feynman rules in CalcHEP format~\cite{Pukhov:1999gg,Belyaev:2012qa}
 (available at the HEPMDB website~\cite{HEPMDB}), upon which
one of two Monte Carlo (MC) partonic event generators was built.
However, since CalcHEP does not implement polarisation, a second code was constructed based on MadGraph~\cite{Stelzer:1994ta}, which agreed entirely with the above one in the unpolarised case. The approximation of our calculations 
is at tree level for what concerns MC event generation whereas some loop corrections were included in the computation 
of the model spectra (again, see~\cite{Barducci:2012kk} for details). 

Finally, although Initial State Radiation (ISR) and Beam-Strahlung (BS) affecting the colliding leptons should in principle be accounted for, we have verified their essentially negligible impact throughout. For the former, the standard expressions of Refs.~\cite{Jadach:1988gb,Skrzypek:1990qs} were considered. Regarding the latter, the parametrisation specified for the ILC project in~\cite{Behnke:2013xla} was adopted: i.e.,
\begin{itemize}
\item[-] beam size $(x+y)$: $645.7$ nm,
\item[-] bunch length: $300$ $\mu$m,
\item[-] bunch population: $2\cdot10^{10}$.
\end{itemize}
Even if the presence of ISR and BS can uniformly affect specific observables (e.g., the asymmetries of the 4DCHM and the SM independently show a uniform relative correction of $5\%$($3\%$) at $\sqrt{s}=370$($500$) GeV), the relative trend of the 4DCHM observables with respect to the SM ones is independent of the presence of ISR and BS. Therefore, for ease of computation and to improve simulation speed, such features will not be considered any further.

\subsection{Results}

Here we will consider three discrete energy values for an $e^+ e^-$ collider, 370, 500 and 1000 GeV. Further, we will
discuss results without and with polarisation of the initial beams in turn. 

\begin{figure}[t!]
\begin{center}
\hspace{-0.8cm}
\vspace{-0.9cm}
\includegraphics[width=0.57\textwidth]{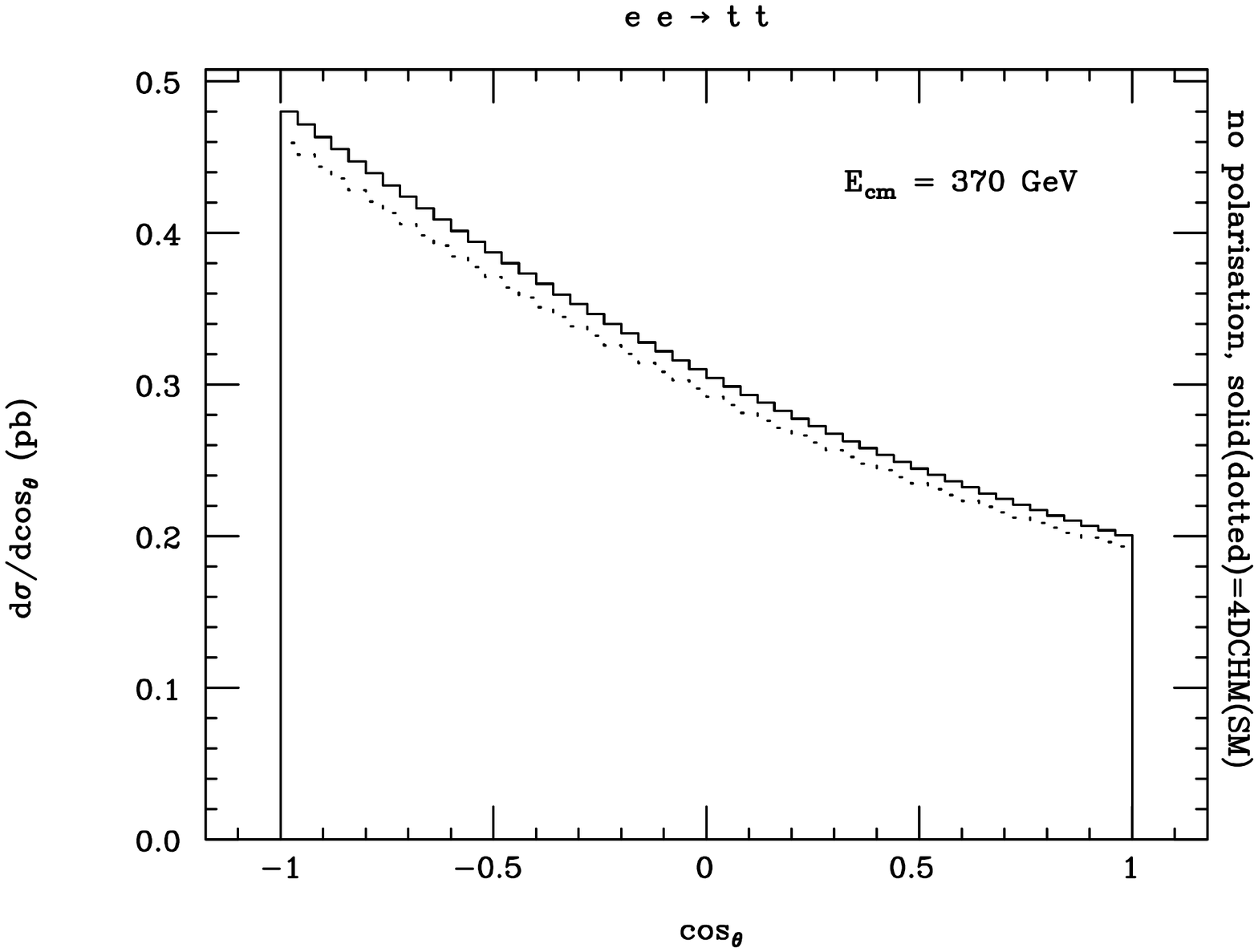}\hspace{-1.6cm}
\includegraphics[width=0.57\textwidth]{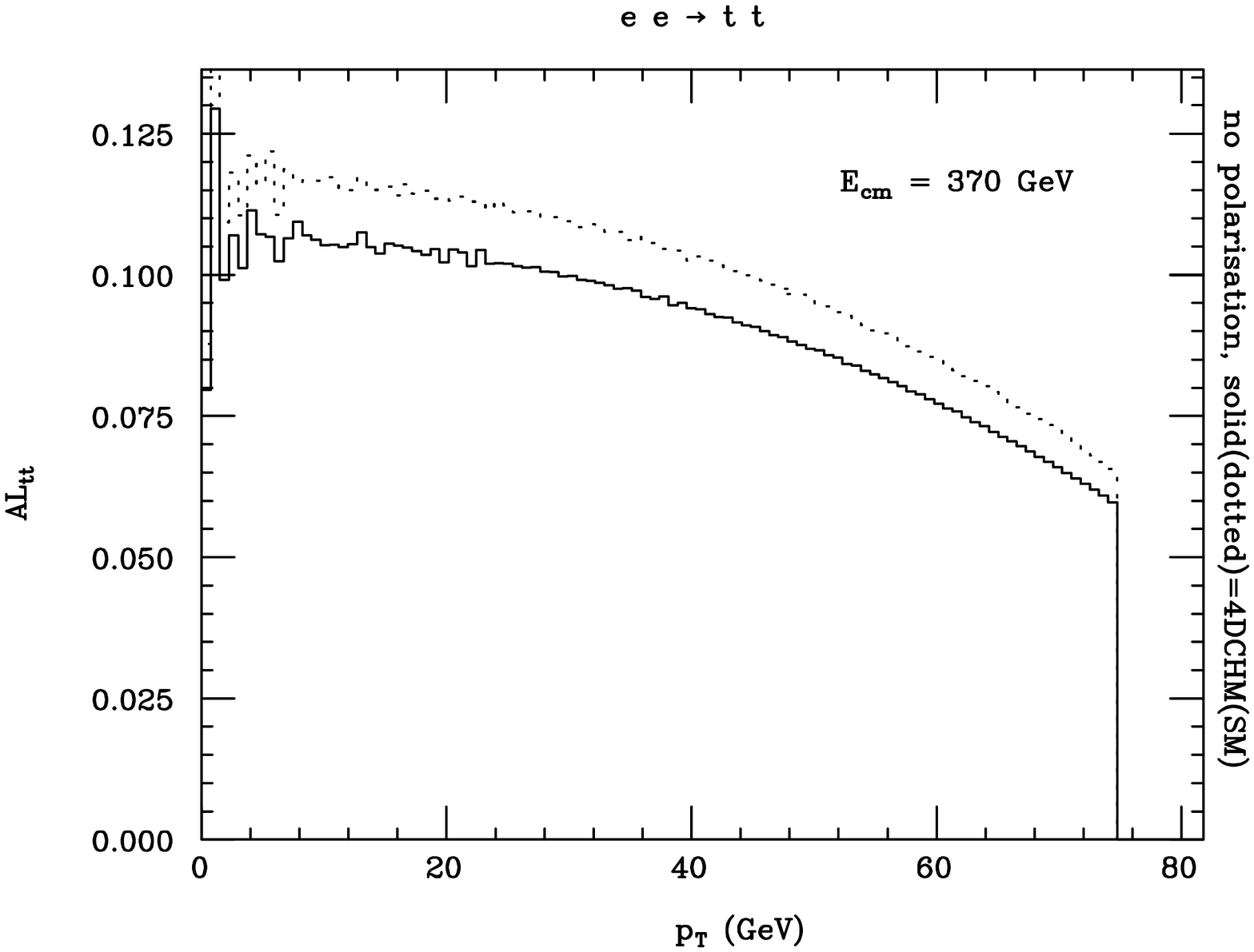}\\
\hspace{-0.8cm}
\vspace{-0.5cm}
\includegraphics[width=0.57\textwidth]{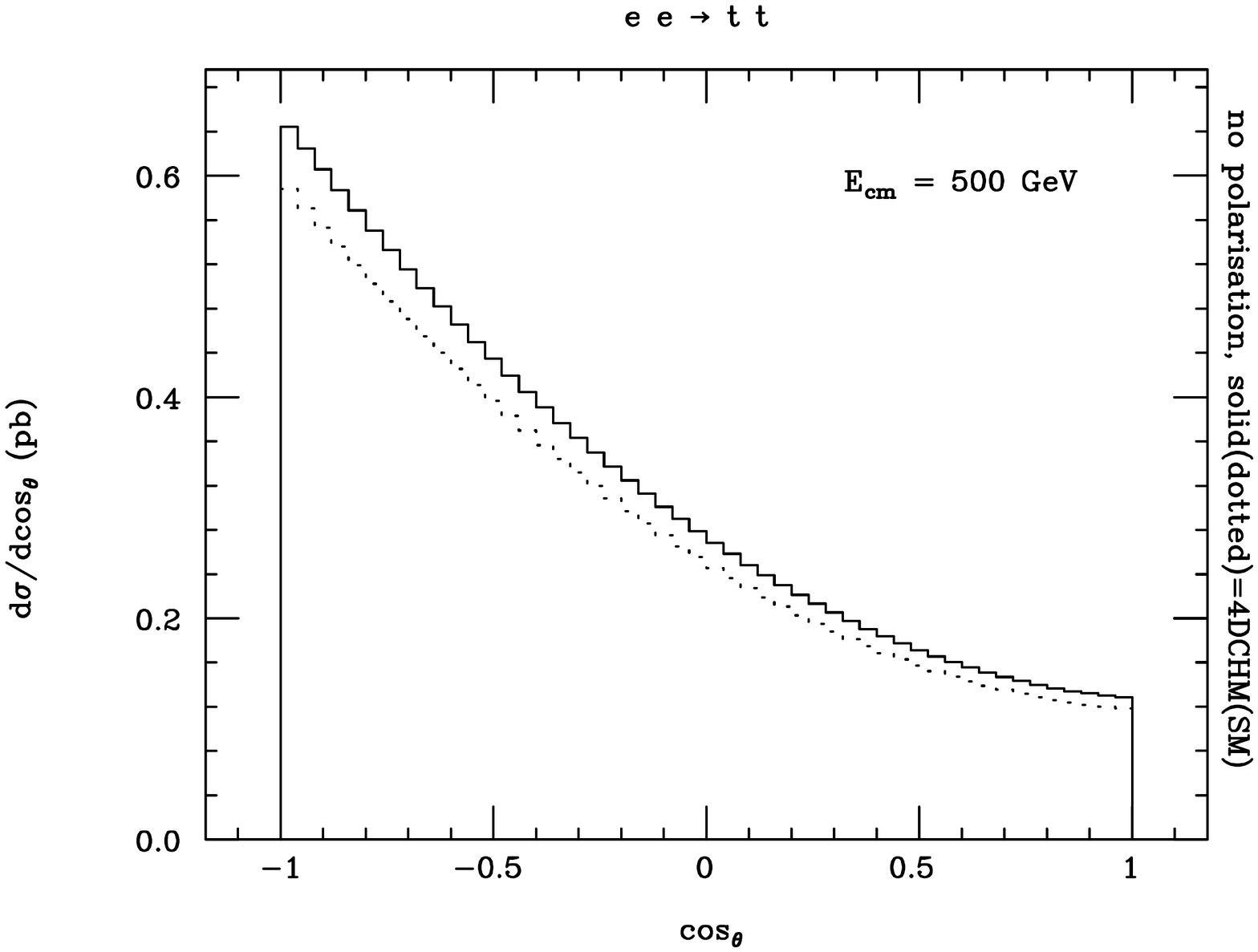}\hspace{-1.6cm}
\includegraphics[width=0.57\textwidth]{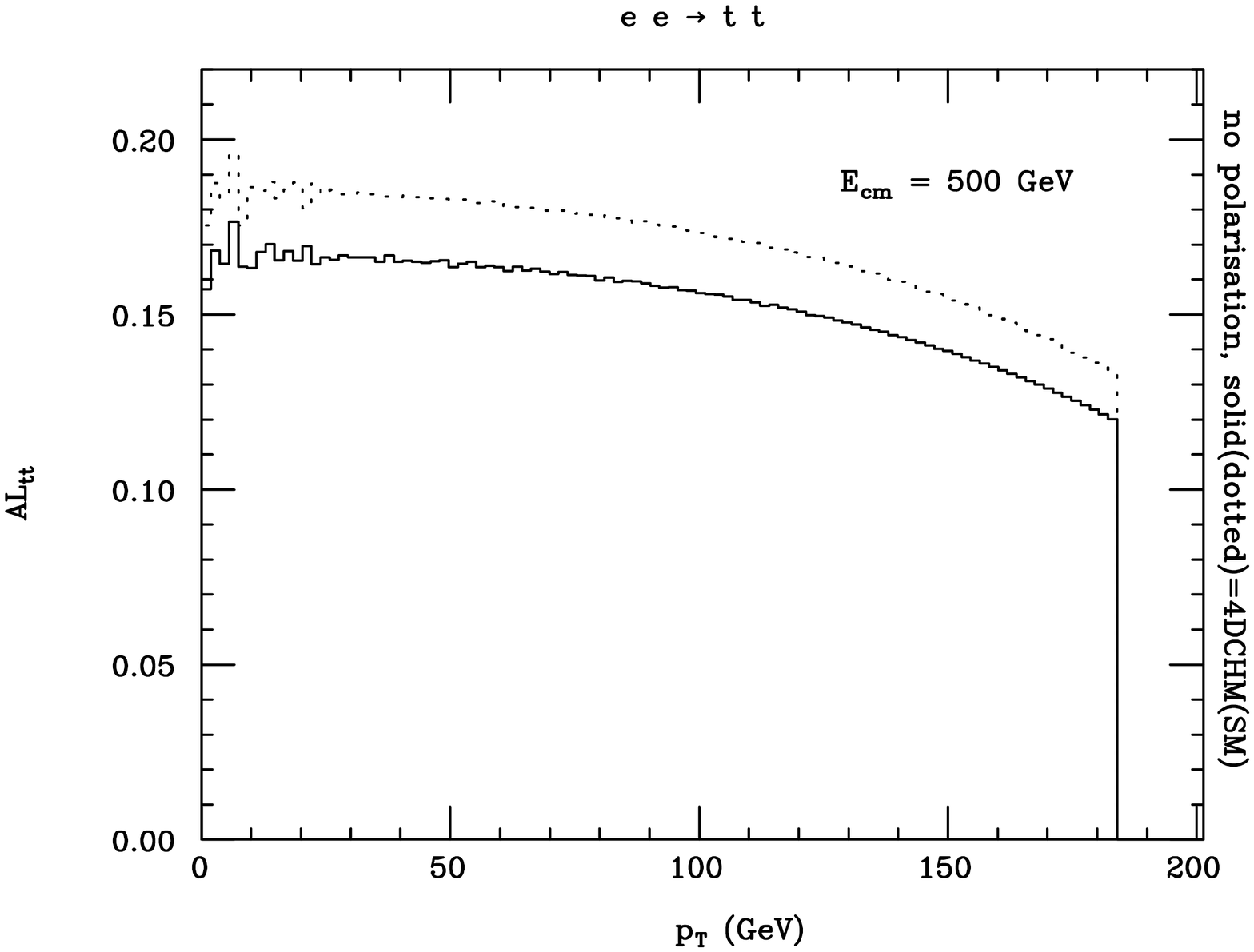}\\
\vspace{-0.5cm}
\hspace{-0.8cm}
\includegraphics[width=0.57\textwidth]{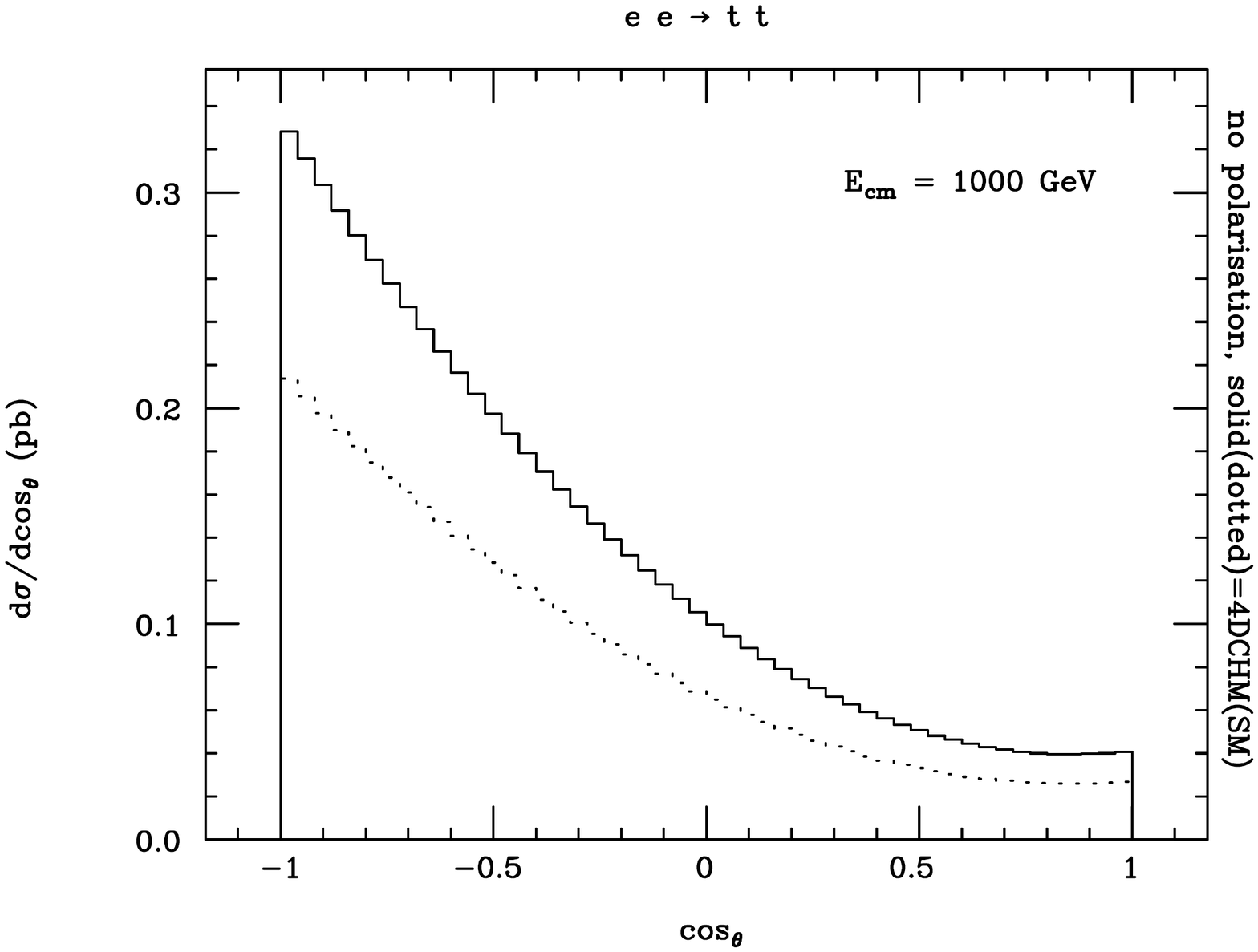}\hspace{-1.6cm}
\includegraphics[width=0.57\textwidth]{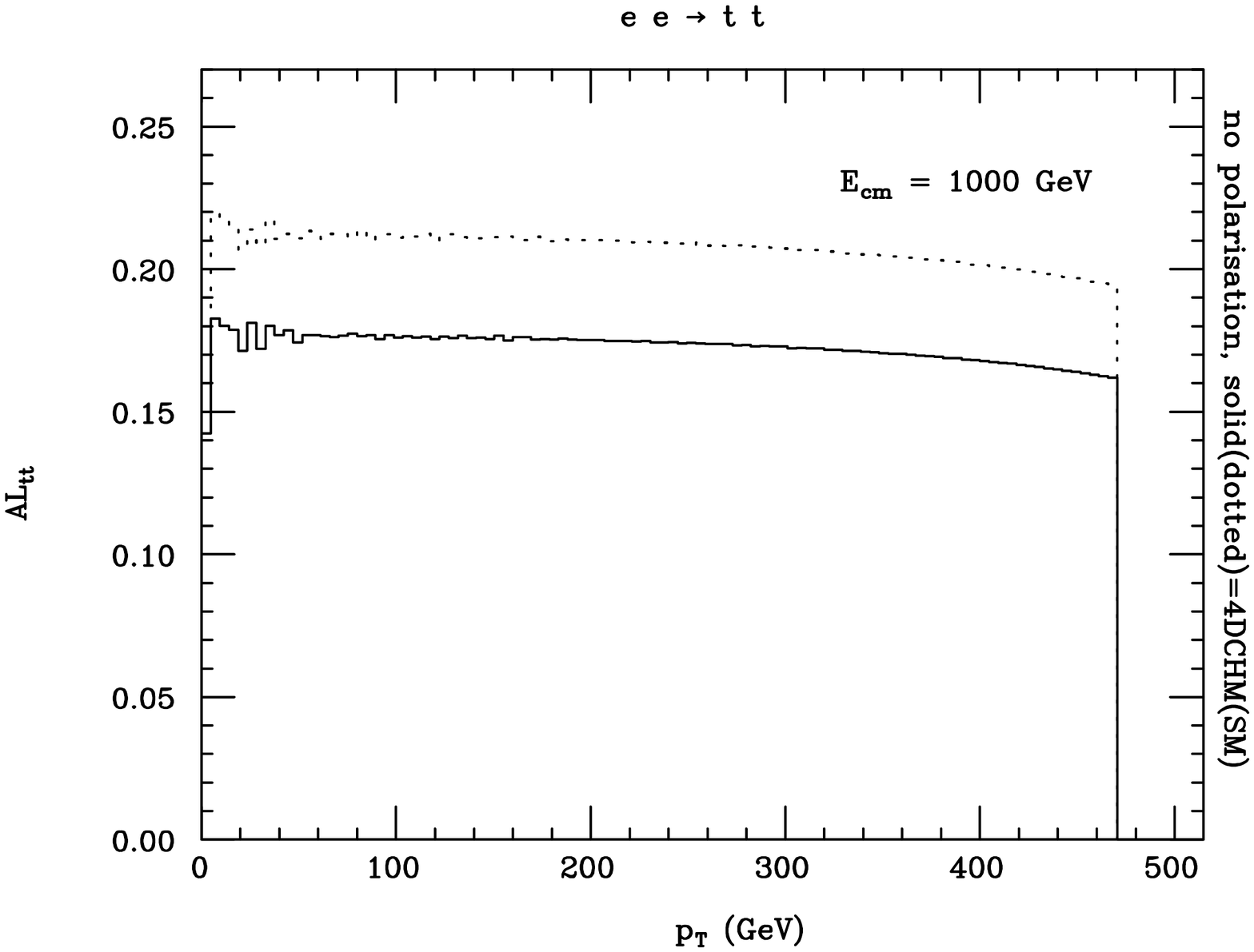}
\vspace{-0.5cm}
\caption{\small \it Differential distributions for the cross section $\sigma$ with respect to $\cos\theta^*$ (left) and for the single spin asymmetry $A_L$ with respect to the $p_T$ (right) of the emitted top quark, for the three different energy options within the SM (dashed) and the 4DCHM (solid) for $M_{Z'_{2,3,5}}$= 2122, 2214, 2831 GeV and $\Gamma_{Z'_{2,3,5}}$= 452, 319, 91 GeV.}
\label{differential}
\end{center}
\end{figure}

To have an idea of the typical deviations induced by the 4DCHM with respect to the SM in the mentioned observables, in Fig.~\ref{differential} we plot, for a single benchmark point ($M_{Z'_{2,3,5}}$= 2122, 2214, 2831 GeV, $\Gamma_{Z'_{2,3,5}}$= 452, 319, 91 GeV), the differential distributions for the cross section $\sigma$ with respect to $\cos\theta^*$ and of the single spin asymmetry $A_L$ with respect to the transverse momentum 
($p_T$) of the emitted top (anti)quark, for the aforementioned different fixed energy options. 
%
We expect that such deviations are all detectable within experimental errors, which are generally claimed to be at the level of percent or even smaller for both the cross section and the asymmetry~\cite{Amjad:2013tlv,Asner:2013hla,Janot:2015yza,Khiem:2015ofa}. Concerning the differential behaviour, the cross section deviations are larger in the backward region while for $A_L$ they are slightly bigger for small $p_T$. Whether or not the differential behaviour can be fully established, also the integrated cross sections and single spin asymmetry do deviate from their SM values: namely, we have, for this particular benchmark point, $|\sigma^{\rm 4DCHM}-\sigma^{\rm SM}|/\sigma^{\rm SM}=4, 9, 53\% $
 and $|A_{L}^{\rm 4DCHM}- A_{L}^{\rm SM}|/A_{L}^{\rm SM}=9, 10, 17\%$ for $\sqrt{s}=370, 500, 1000$ GeV, respectively. Such dynamics at differential level is very typical over a wide collection of kinematic observables and the fact that we have chosen here $\sigma$ and $A_L$ as reference measures is not coincidental, as we shall see that they are affording the largest corrections.

But let us now concentrate on the integrated values of the cross sections and asymmetries in order to disentangle the various sources of deviations with respect to the SM expectations. In doing this exercise, we do not enforce selection cuts,
as we are working with on-shell top quarks whereas these are applied to their decay products. However, we do not
expect that finite efficiencies due to enforcement of selection cuts will affect our conclusions. 
\begin{figure}[t!]
\begin{center}
\hspace{-0.5cm}
\includegraphics[width=0.33\textwidth]{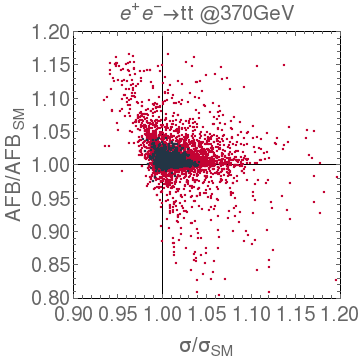}
\includegraphics[width=0.33\textwidth]{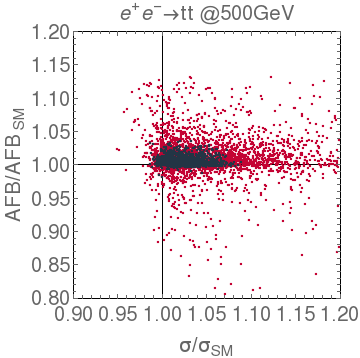}
\includegraphics[width=0.33\textwidth]{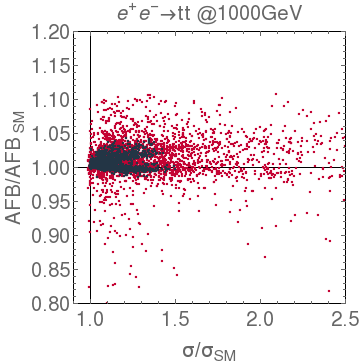}\\\hspace{-0.3cm}
\includegraphics[width=0.33\textwidth]{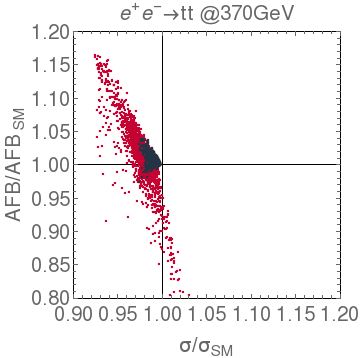}
\includegraphics[width=0.33\textwidth]{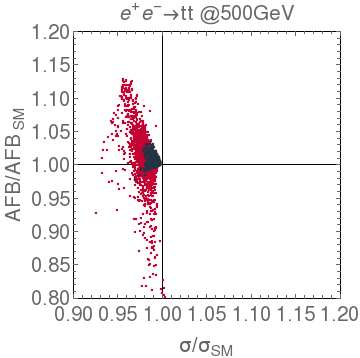}
\includegraphics[width=0.33\textwidth]{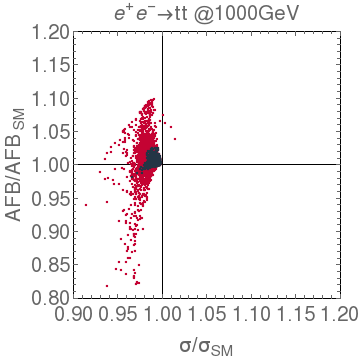}
\vspace{-0.4cm}
\caption{\small \it Predicted deviations for the cross section versus $A_{FB}$ for the process $e^+e^-\to t\bar t$ at 370, 500, 1000 GeV in the 4DCHM compared with the SM (top panel) and the corresponding ones with removed $Z'$ exchange in the $s$-channel (bottom panel). The points correspond to $f=0.75-1.5$ TeV, $g_\rho=1.5-3$. The colour code is the same of Fig.~\ref{tLR}. }
\label{xs_AFB}
\end{center}
\end{figure}

The results of the aforementioned scan mapped in $\sigma$ and $A_{FB}$ for the three customary choice of the
$e^+e^-$ CM energy are found in Fig.~\ref{xs_AFB}. Herein, we can appreciate the importance of the interference between the SM gauge bosons and the $Z'$s. In fact, the shown correlations between the expected deviations in 
$\sigma$ and $A_{FB}$ are dramatically different depending on whether we include or not the propagation of the
$Z'$ states, especially for the cross section.
The effect of $Z'$ exchange in the cross section is very important already at $\sqrt{s}$=370 GeV. In fact, the interference tends to compensate the lowering of the cross section due to the coupling modification (see bottom panels of Fig.~\ref{xs_AFB}) and finally gives a positive contribution which grows with energy, just like the interference does, when approaching the mass value of the new resonances (as previously said, there are mainly two nearly-degenerate $Z'$s contributing to the process, namely $Z'_2$ and $Z'_3$). Deviations up to 50\% are expected in the total cross section at $\sqrt{s}$=1000 GeV while the effect on the $A_{FB}$ is less evident. The relatively different effect is understood, as for this observable we are dividing by the total cross section and this washes out the large $Z'$ interference dependence,
which is similar in both the forward and backward hemispheres. The overall size of the deviations does not change when we enforce a larger cut of 1 TeV on the mass of the extra fermions, thus assuming that no extra fermions will be observed at LHC Run 2 with mass larger than 1 TeV. Anyhow, this stronger mass cut slightly reduces the deviations on $A_{FB}$
 but it affects $\sigma$ very little.

\begin{figure}[!t]
\begin{center}
\hspace{-0.5cm}
\includegraphics[width=0.33\textwidth]{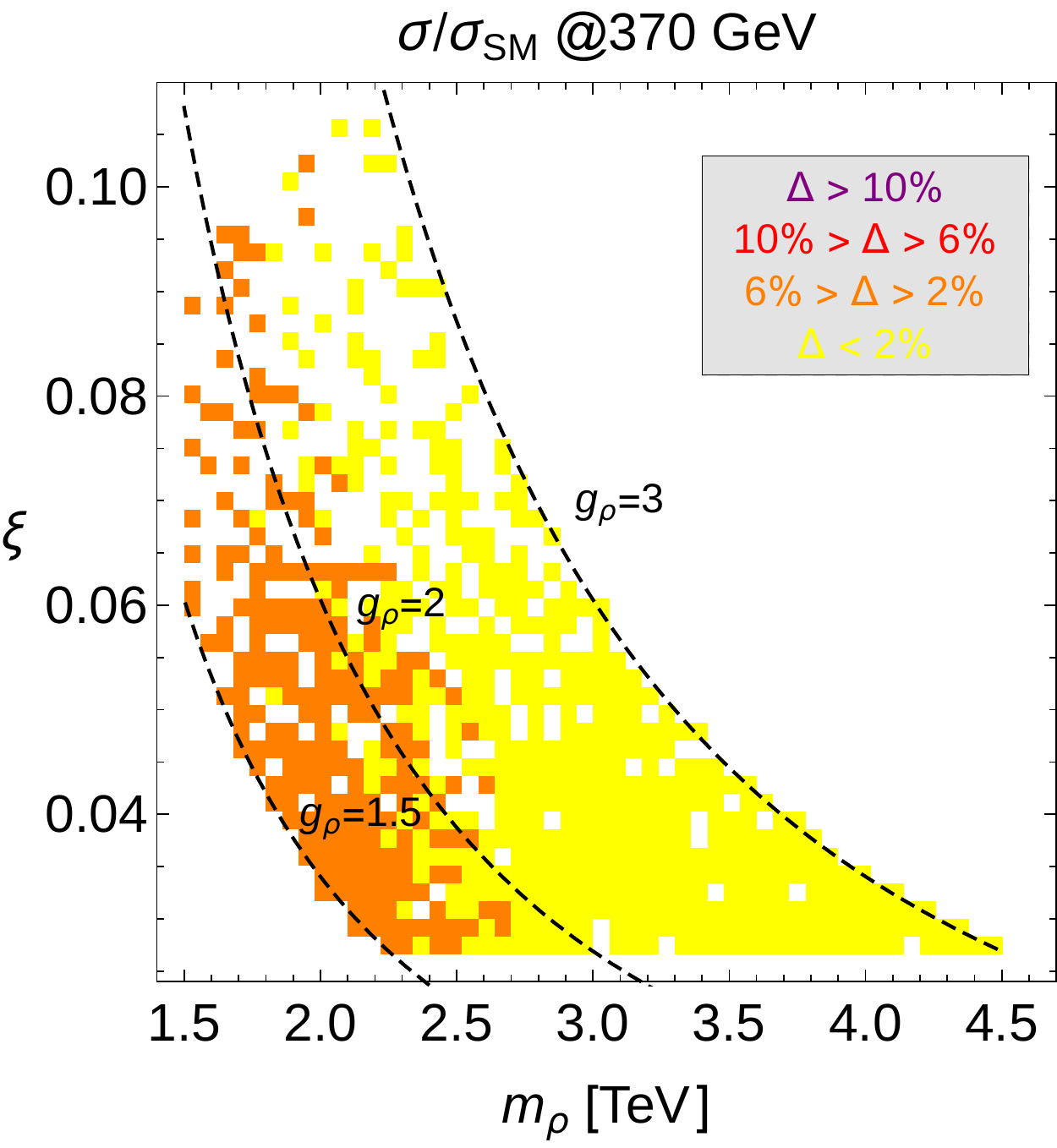}
\includegraphics[width=0.33\textwidth]{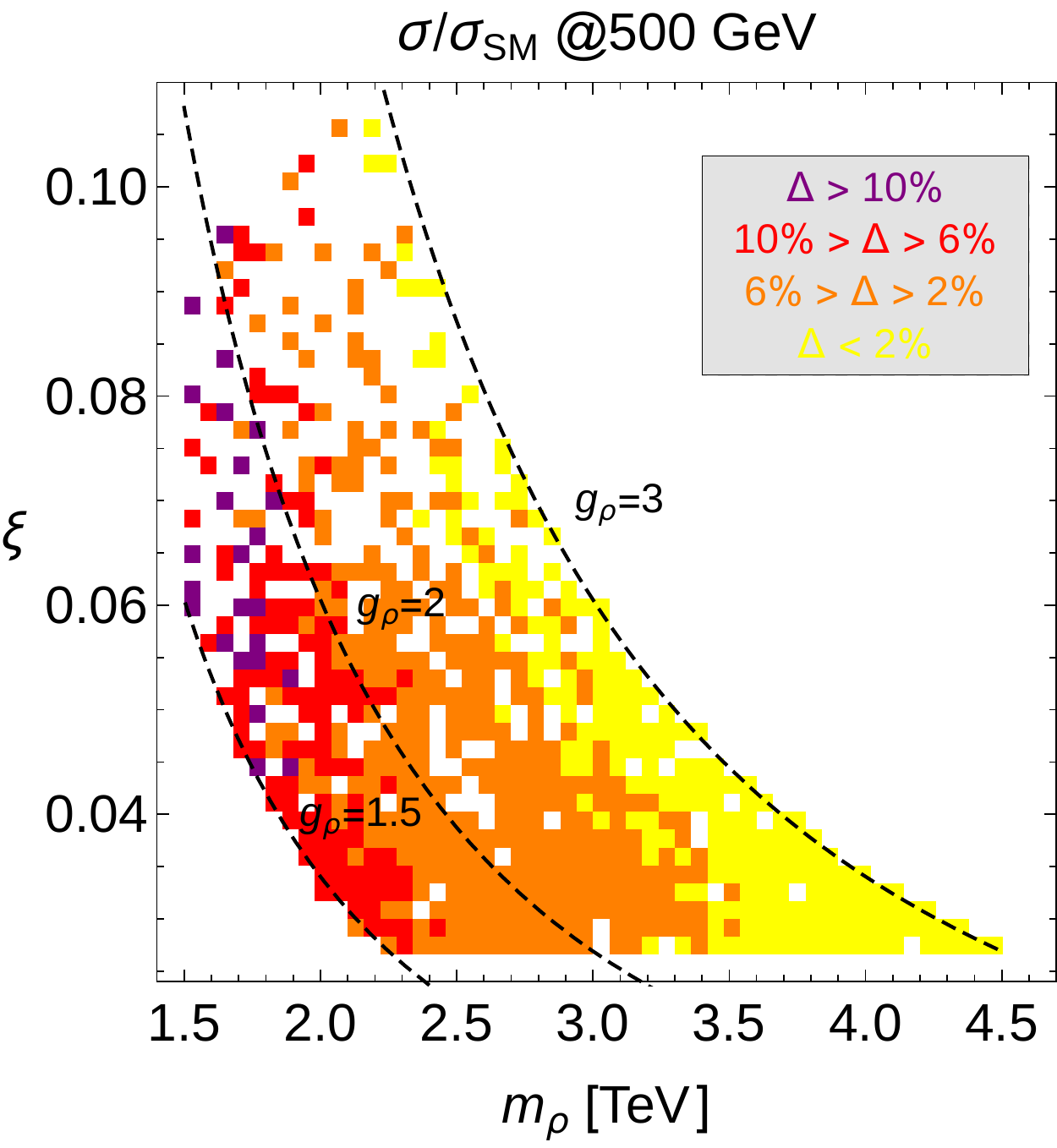}
\includegraphics[width=0.33\textwidth]{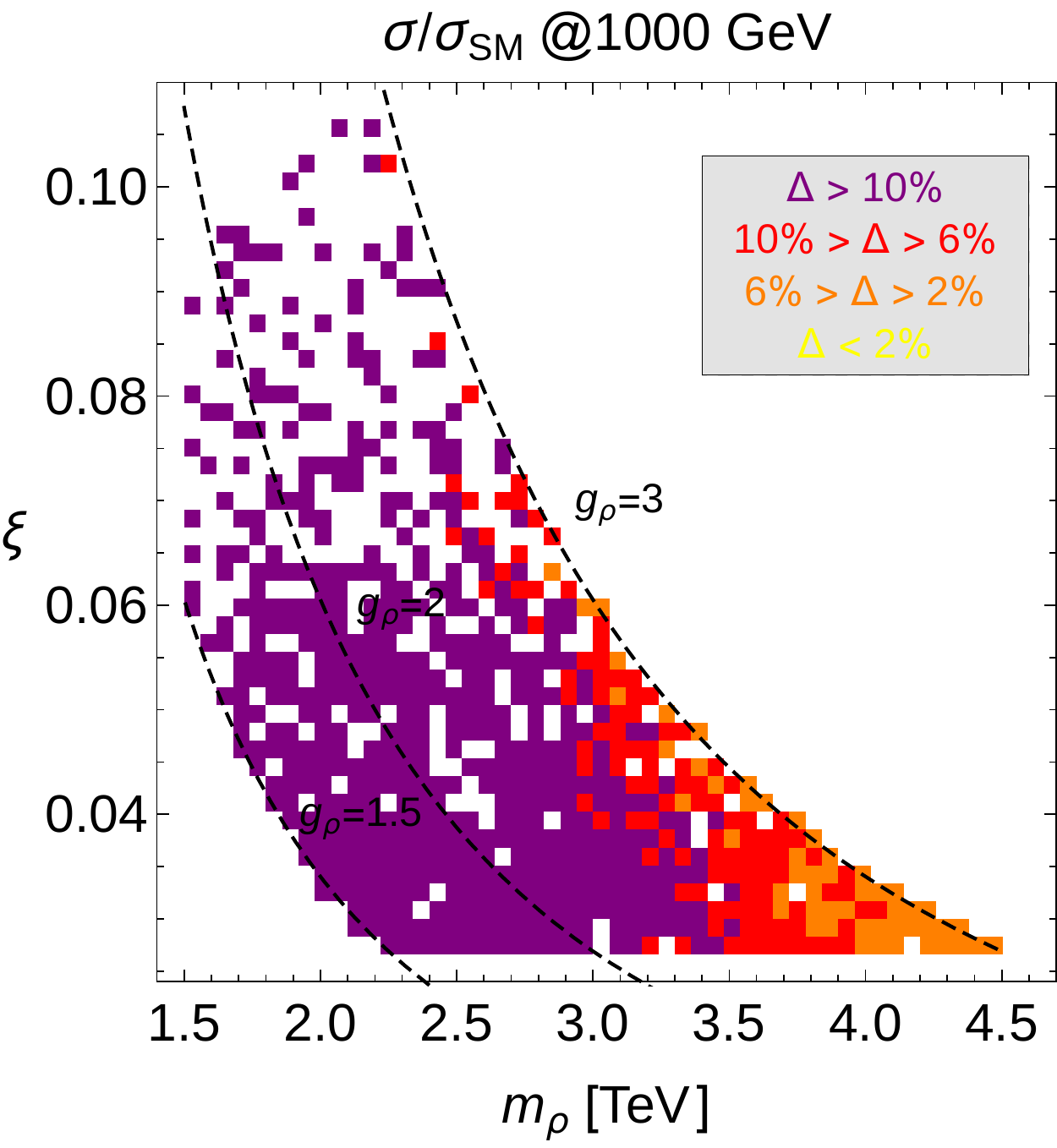}\\
\vspace{-0.1cm}
\caption{\small \it Predicted deviations for the cross section of the process $e^+e^-\to t\bar t$ at 370, 500, 1000 GeV in the 4DCHM compared with the SM as functions of $m_\rho=f g_\rho$ and $\xi=v^2/f^2$. 
For each point we have selected the configuration yielding the maximal deviation defined as $\Delta= (\sigma^{\rm 4DCHM}- \sigma^{\rm SM})/\sigma^{\rm SM}$. The points correspond to $f=0.75-1.5$ TeV, $g_\rho=1.5-3$. Bounds on the masses of the extra fermions are the same as in Fig.~\ref{tLR}.}
\label{colour_xi}
\end{center}
\end{figure}

 Let us now extract the sensitivity to the relevant parameters of a typical CHM, herein realised via the 4DCHM, from the various $e^+e^-\to t \bar t$ observables. In Fig.~\ref{colour_xi} we plot, by using different colours, the predicted deviations for the cross section at $\sqrt{s}$= 370, 500, 1000 GeV in the 4DCHM compared with the SM as functions of $m_\rho=f g_\rho$, the typical mass of the $Z'$s up to EW corrections, and $\xi=v^2/f^2$, the compositeness parameter.
For each point we have selected the configuration yielding the maximal deviation defined as $\Delta= (\sigma^{\rm 4DCHM}- \sigma^{\rm SM})/\sigma^{\rm SM}$. The points correspond to $f=0.75-1.5$ TeV, $g_\rho=1.5-3$.
We see that, by requiring a deviation larger than 2\% to be detected, a 500 GeV machine is sensitive to new spin-1 resonances with mass up to 3.5 TeV. Of course there could be a configuration of fermion parameters such that the width of the $Z'$ is sufficiently large to give a smaller deviation. In this respect the ones in Fig.~\ref{colour_xi} must be interpreted as the maximal sensitivities for each given CM energy. However, by combining different observables, the reach in mass would improve no matter the actual width.

\begin{figure}[t!]
\begin{center}
\hspace{-0.5cm}
\includegraphics[width=0.33\textwidth]{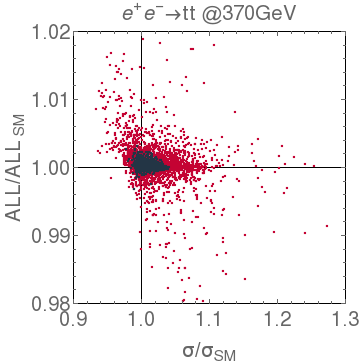}
\includegraphics[width=0.33\textwidth]{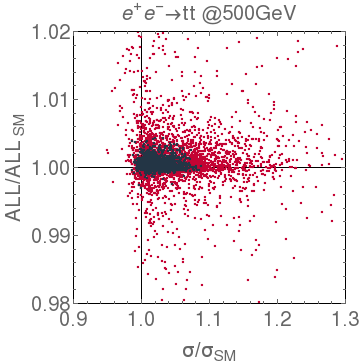}
\includegraphics[width=0.33\textwidth]{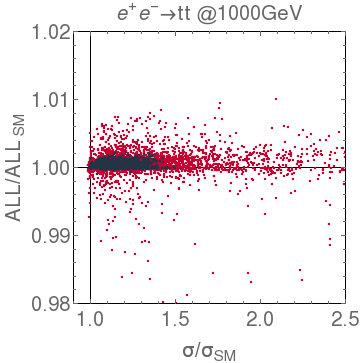}\\\hspace{-0.3cm}
\includegraphics[width=0.33\textwidth]{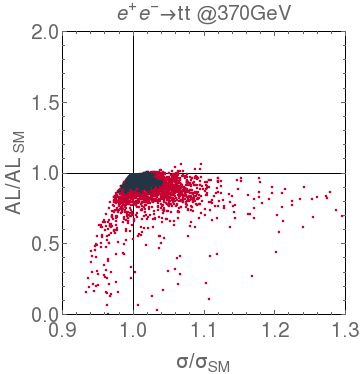}
\includegraphics[width=0.33\textwidth]{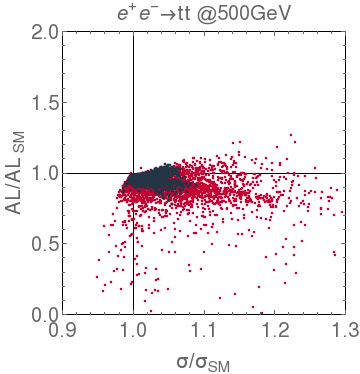}
\includegraphics[width=0.33\textwidth]{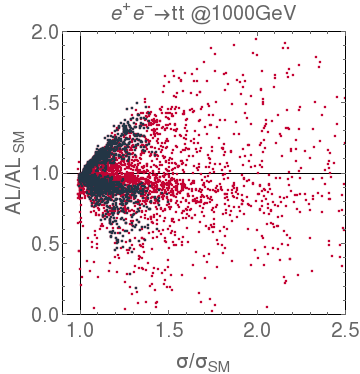}
\vspace{-0.1cm}
\caption{\small \it Predicted deviations for the cross section versus $A_{LL}$ (top) and $A_{L}$ (bottom) for the process $e^+e^-\to t\bar t$ at 370, 500, 1000 GeV in the 4DCHM compared with the SM. The points correspond to $f=0.75-1.5$ TeV, $g_\rho=1.5-3$. The colour code is the same of Fig.~\ref{tLR}. }
\label{xs_ALL}
\end{center}
\end{figure}

Concerning the spin asymmetries, $A_L$ deserves particular attention. In fact, while the double spin asymmetry $A_{LL}$ shows the same behaviour as $A_{FB}$ albeit with smaller deviations (see Fig.~\ref{xs_ALL} (top)), the single spin asymmetry $A_L$ is sensitive to the relative sign of the left- and right-handed couplings of the $Z$ and $Z'$s to the top pair. It is unique in offering the chance to separate $Z'_2$ and $Z'_3$ as the two 4DCHM objects contributing to this asymmetry in opposite directions. In Fig.~\ref{xs_ALL} (bottom) the expected affects are shown. Notice that, for $\sqrt{s}=1000$ GeV, where the interferences of $Z'_2$ and $Z'_3$ with $\gamma,Z$ are largest, the two contributions to $A_L$ appear visible in opposite directions and the deviations can reach 50\%.

This effect is emphasised if electron and positron beam polarisations are available. In fact, the $Z'_2$ and $Z'_3$ interferences have opposite sign. For a possible beam configuration in presence of polarisation, 
we use the following definition, according to~\cite{MoortgatPick:2005cw}:
\begin{equation}
\sigma_{P,P'}=\frac 1 4[(1- P P')(\sigma_{-,+}+\sigma_{+,-}+(P-P ')(\sigma_{+,-}-\sigma_{-,+})]
\end{equation}
with $\sigma_{-,+(+,-)}=\sigma(e^-_L,e^+_R)(\sigma(e^-_R,e^+_L))$ and $P(P')$ the polarisation degree for electrons(po\-si\-trons). Similar expressions hold for the asymmetries. 

\begin{figure}[t!]
\begin{center}
\hspace{-0.5cm}
\includegraphics[width=0.33\textwidth]{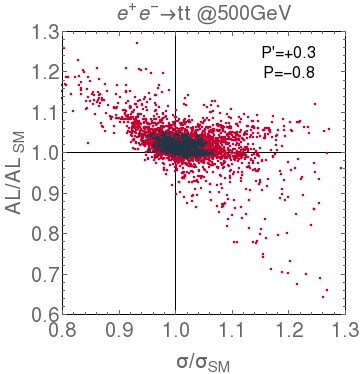}
\includegraphics[width=0.33\textwidth]{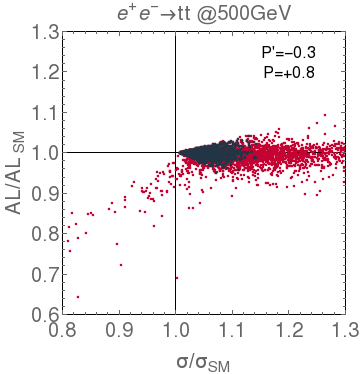}
\includegraphics[width=0.33\textwidth]{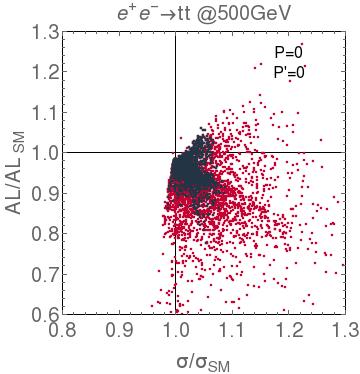}
\vspace{-0.1cm}
\caption{\small \it Predicted deviations for the cross section versus $A_{L}$ for the process $e^+e^-\to t\bar t$ at 500 GeV in the 4DCHM compared with the SM for $P'(e^+)=+0.3$ and $P(e^-)=-0.8$ (left), $P'(e^+)=-0.3$ and $P(e^-)=+0.8$ (center) and for unpolarised beams (right). The points correspond to $f=0.75-1.5$ TeV, $g_\rho=1.5-3$. The colour code is the same of Fig.~\ref{tLR}. }
\label{xs_AL_pol}
\end{center}
\end{figure}

In Fig.~\ref{xs_AL_pol} the correlated deviations for $\sigma$ and $A_L$ at $\sqrt{s}$=500 GeV for $P'(e^+)=+0.3$ and $P(e^-)=-0.8$ (left), $P'(e^+)=-0.3$ and $P(e^-)=+0.8$ (center) and for unpolarised beams (right) are shown. 
Corrections to the cross section are slightly larger in the case of polarised beams when negative whereas they are similar when positive.
Somewhat unintuitively, for the case of the asymmetry, they are largest for the unpolarised case when negative whereas they are comparable when positive. This can be understood by recalling that an initial state polarisation does not automatically select one in the final state, as all helicity combinations 
 in the latter are always possible whichever is enforced in the former. Furthermore, for the asymmetry, one should be reminded of the fact that this observable is normalised to the total cross section. Therefore, it is apparent that the potential benefits of the
corrections seen in Fig.~2, wherein 4DCHM effects are largest on the chiral coefficient of the top quark which is smallest, 
combined with the fact that $s$-channel exchange of $Z'$s becomes dominant with increasing energy, 
does not translate into a preferred beam configuration to be adopted in order to highlight the 4DCHM effects. 
Results are shown here for 500 GeV, but the pattern remains similar ar both lower and higher CM energies. 
However, as we shall see in the next Section, beam polarisation plays a key role in disentangling the presence of two
competing effects, due to the $Z'_2$ and $Z'_3$ states, which largely cancel in the unpolarised case.

\section{\label{top-selection} Disentangling the 4DCHM effects}

\begin{figure}[!t]
\begin{center}
\hspace{-0.5cm}
\includegraphics[width=0.5\textwidth]{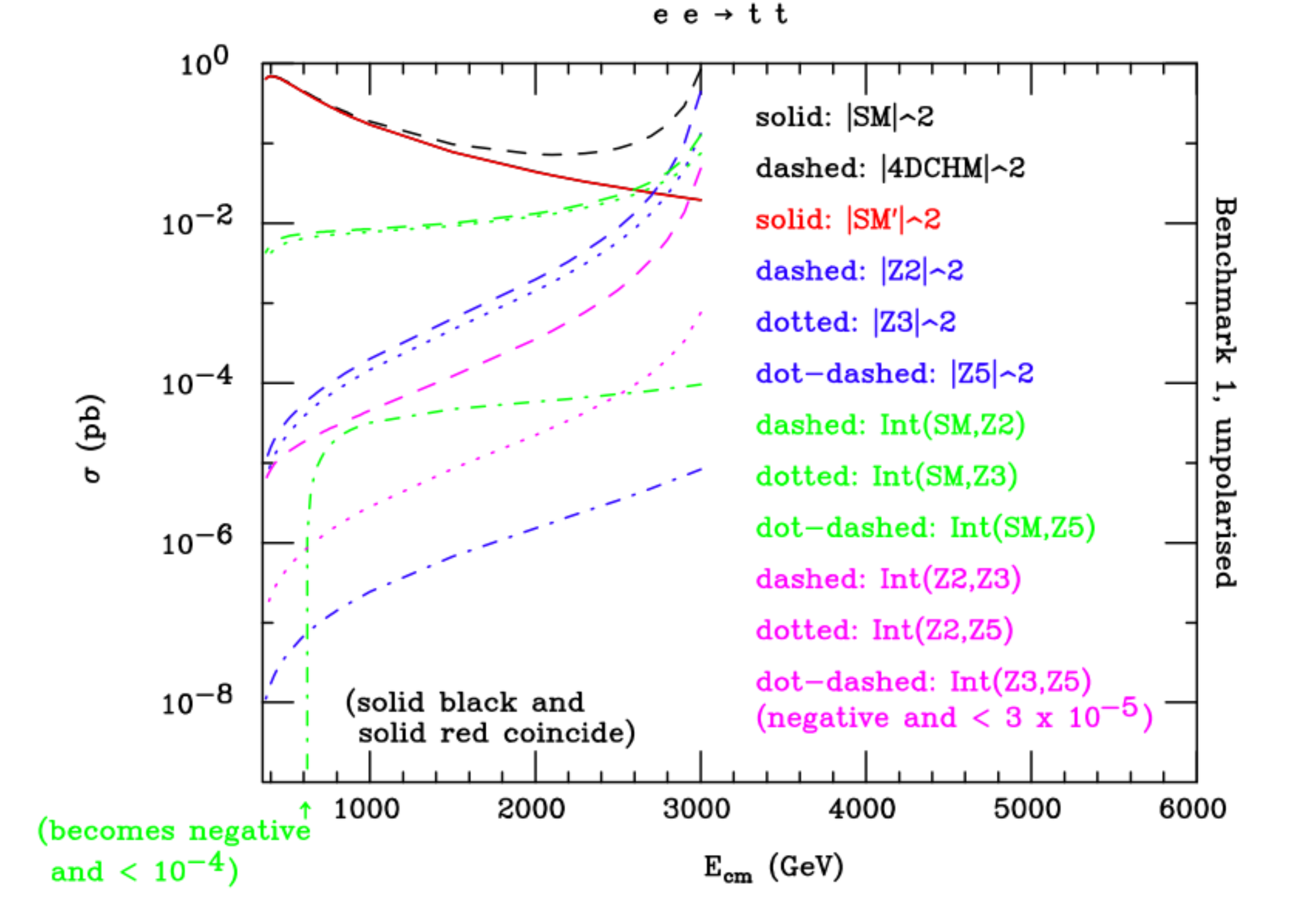}
\includegraphics[width=0.5\textwidth]{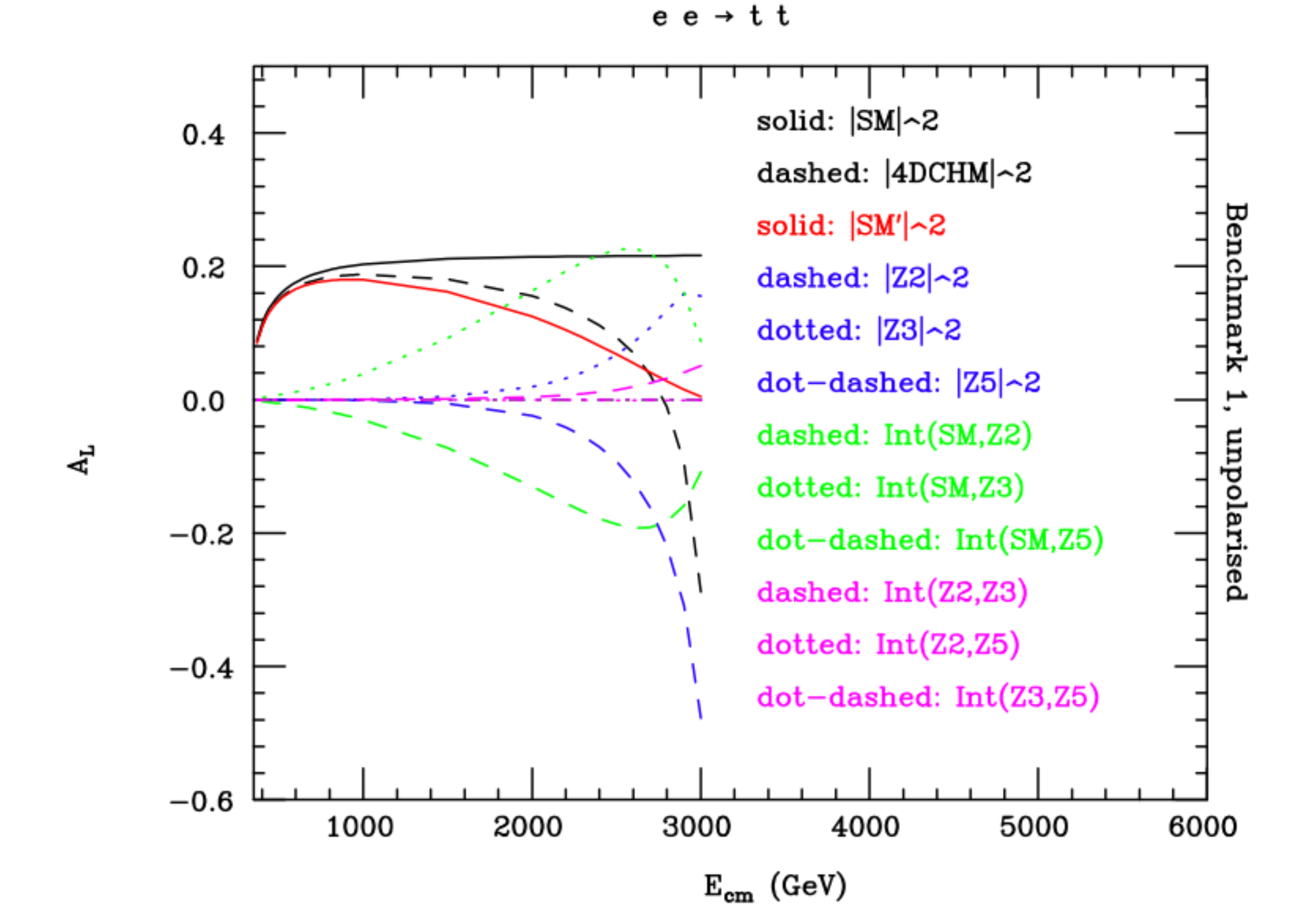}
\caption{\small \it Contributions to the unpolarised cross section $\sigma$ and single spin asymmetry $A_L$ as functions of $\sqrt{s}$ for the SM and a 4DCHM benchmark point corresponding to $M_{Z'_{2,3,5}}= 3087, 3143, 4252$ GeV and $\Gamma_{Z'_{2,3,5}}= 53, 85, 90$ GeV.
}
\label{split}
\end{center}
\end{figure}

It is interesting to disentangle the various effects onsetting the 4DCHM deviations. 
In Fig.~\ref{split} we plot the various contributions to the $\sigma$ and to $A_L$ for a particular benchmark point with $M_{Z'_{2,3,5}}= 3087, 3143, 4252$ GeV and $\Gamma_{Z'_{2,3,5}}= 53, 85, 90$ GeV, as function of $\sqrt{s}$ (now up to CLIC energies, to emphasise the presence of the poles due to the relevant $Z'$ states in $s$-channel).
We call $|{\rm SM}|^2$($|{\rm 4DCHM}|^2$) the full SM(4DCHM) results, while we split the NP contributions as follows:
\begin{enumerate}
\item $|{\rm SM}'|^2$ is due to the square of the $\gamma,Z$ diagrams of the SM with the couplings rescaled;
\item $|Z'_2|^2$ is due to the square of the $Z'_2$ diagram;
\item $|Z'_3|^2$ is due to the square of the $Z'_3$ diagram;
\item $|Z'_5|^2$ is due to the square of the $Z'_5$ diagram;
\item Int(SM,$Z'_2$) is due to the interference between the diagrams in 1 and 2;
\item Int(SM,$Z'_3$) is due to the interference between the diagrams in 1 and 3;
\item Int(SM,$Z'_5$) is due to the interference between the diagrams in 1 and 4;
\item Int($Z'_2,Z'_3$) is due to the interference between the diagrams in 2 and 3;
\item Int($Z'_2,Z'_5$) is due to the interference between the diagrams in 2 and 4;
\item Int($Z'_3,Z'_5$) is due to the interference between the diagrams in 3 and 4.
\end{enumerate}
(Notice that 1 to 10 sum to $|{\rm 4DCHM}|^2$.)

\begin{figure}[!t]
\begin{center}
\hspace{-0.5cm}
\includegraphics[width=0.45\textwidth]{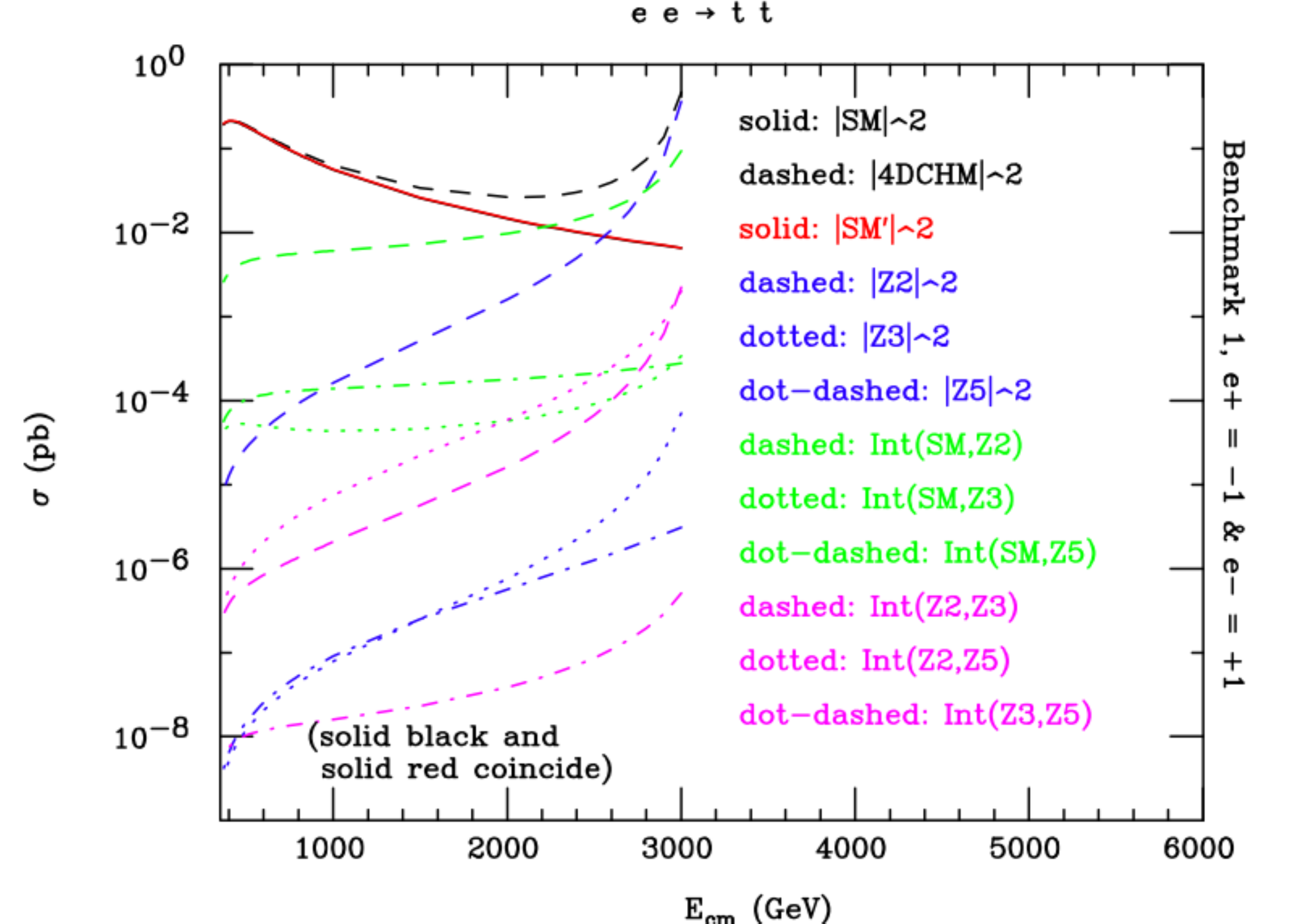}
\includegraphics[width=0.45\textwidth]{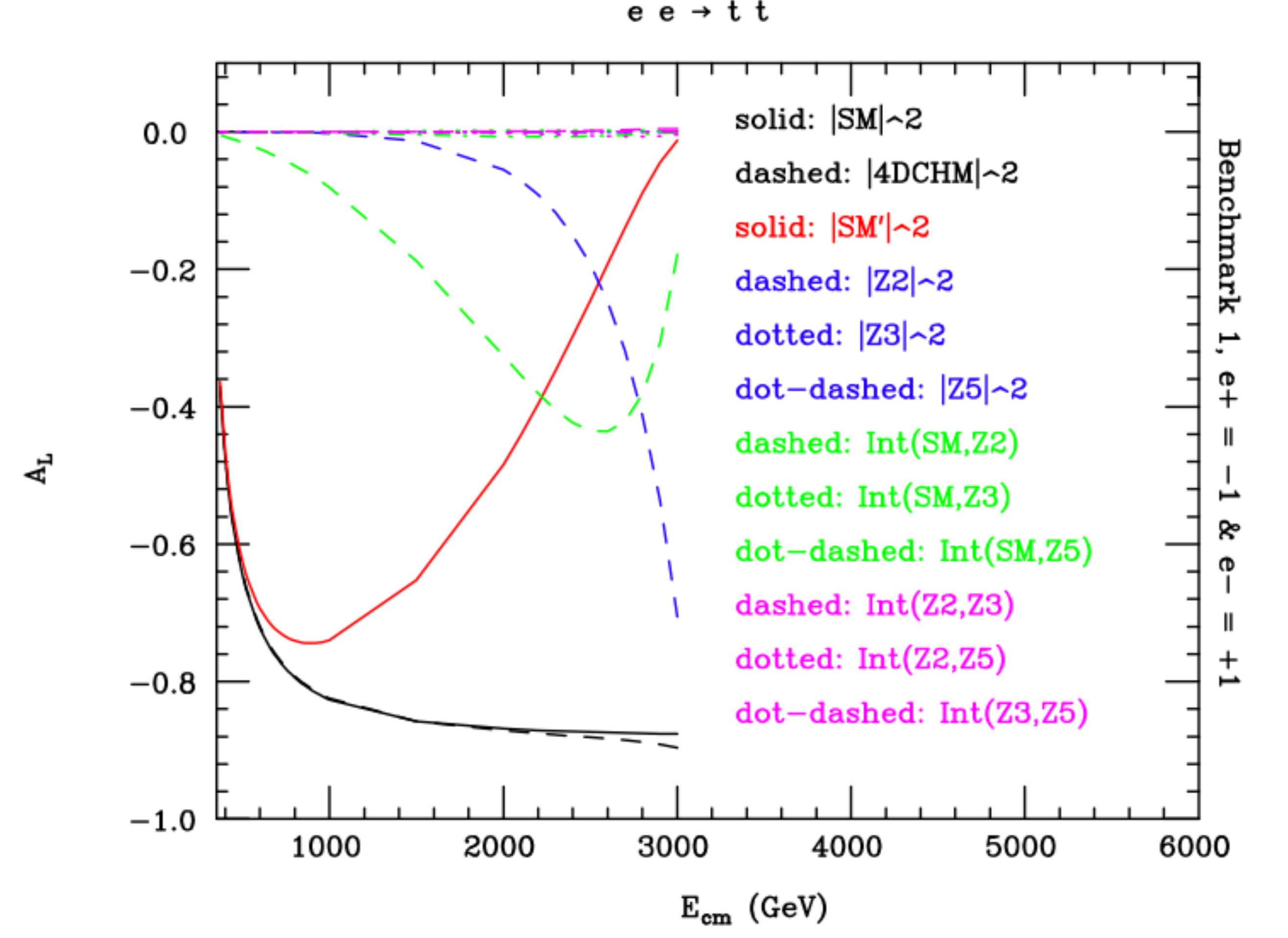}\\
\hspace{-0.5cm}\vspace{-0.6cm}
\includegraphics[width=0.45\textwidth]{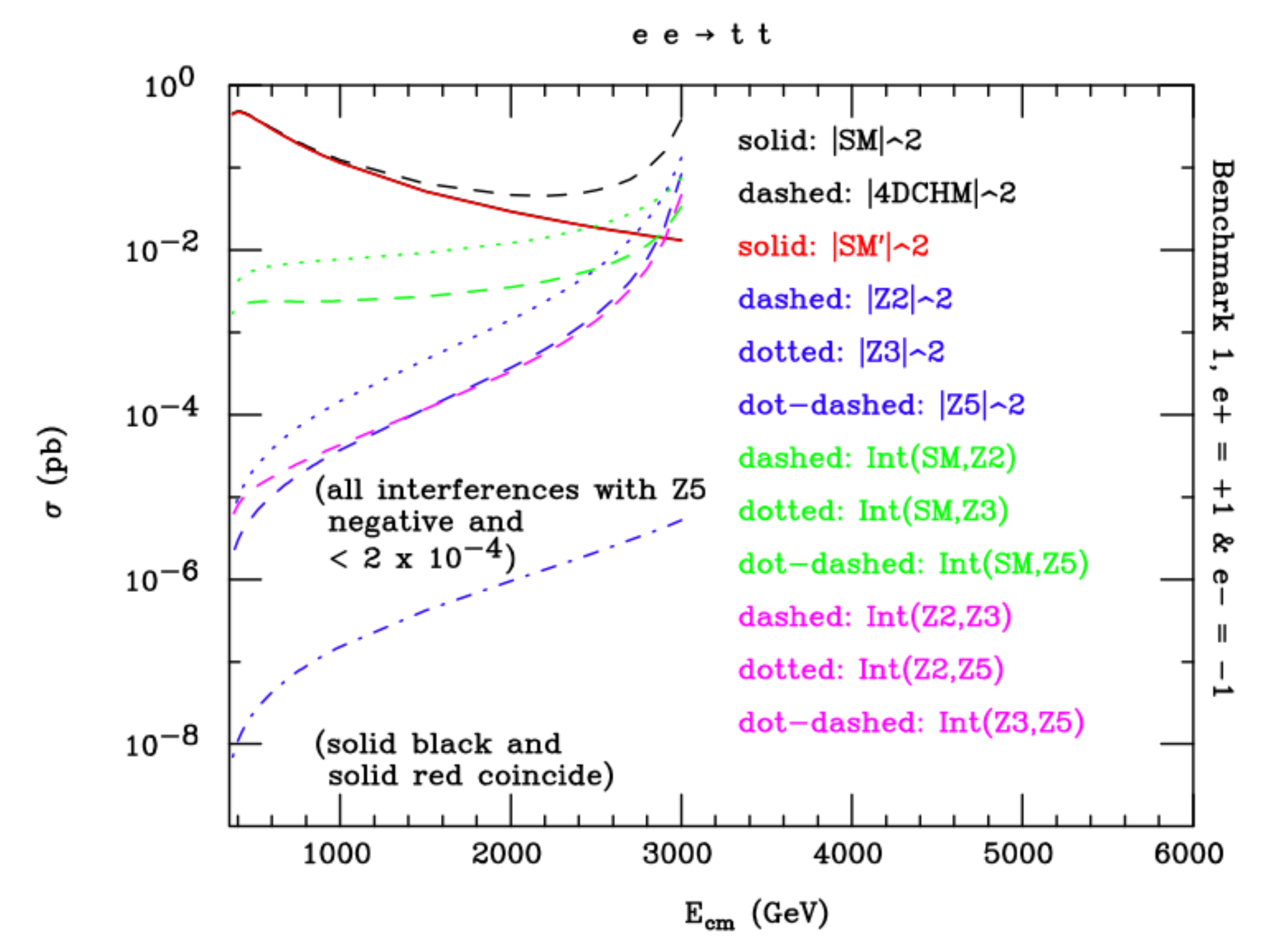}
\includegraphics[width=0.45\textwidth]{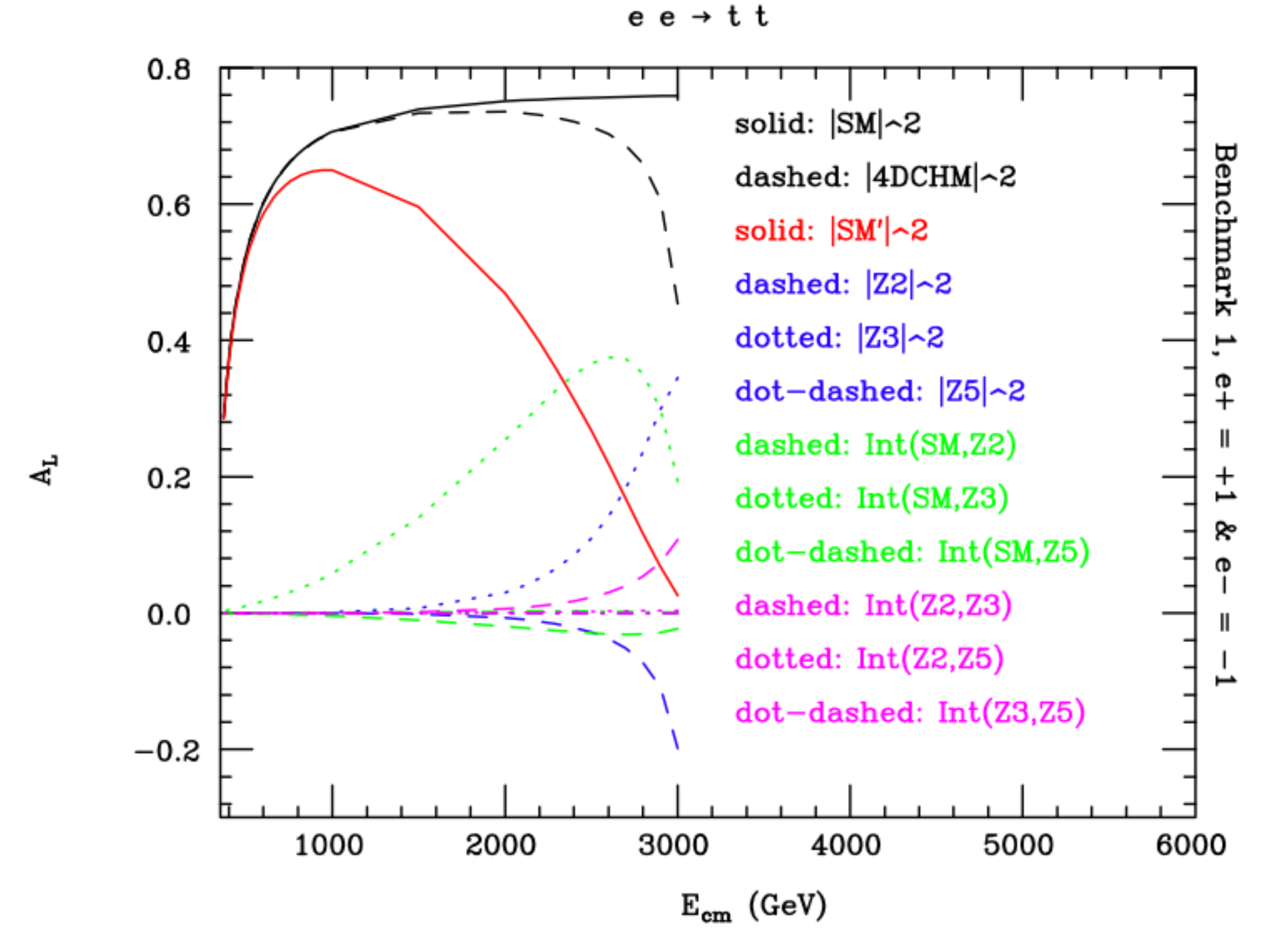}\\[0.5cm]
\caption{\small \it Contributions to the polarised cross section $\sigma$ and single spin asymmetry $A_L$ as functions of $\sqrt{s}$ for the SM and a 4DCHM benchmark point corresponding to $M_{Z'_{2,3,5}}= 3087, 3143, 4252$ GeV and $\Gamma_{Z'_{2,3,5}}= 53, 85, 90$ GeV. Here $ P=1$, $P'=-1$ (top) and $P=-1$, $P'=1$ (bottom).
}
\label{splitpol}
\end{center}
\end{figure}

It is quite evident the importance of the interference between $Z'_{2,3}$ and the SM gauge bosons, as 
they are always amongst the largest contributors to the total 4DCHM cross section (left plot of Fig.~\ref{split}), only
second to the rescaled SM contribution (at low CM energy) or the contributions due to the $Z'_{2,3}$ resonances (at high
CM energy). What is most remarkable (right plot of Fig.~\ref{split}) is that these two contributions tend to cancel in
the single spin asymmetry, a reflection of the opposite signs and similar strength 
 that the $Z'_{2,3}tt$ vertices have in the two different chiral coefficients.
%
To render this manifest, 
for the same benchmark point, we have studied the effects of initial beam polarisation. In Fig.~\ref{splitpol} the two extreme cases $P=1$, $P'=-1$ and $P=-1$, $P'=1$ are considered. While these will not be achieved experimentally, they are useful in order to single out the 4DCHM effects. After all, the various components of the
matrix element plotted in Figs.~\ref{split}--\ref{splitpol} (except $|{\rm SM}|^2$ and $|{\rm 4DCHM}|^2$)
 have no meaning per se, as they cannot be separated experimentally either. Anyhow, the two plots
in the figure make evident the aforementioned features of the $Z'_{2,3}tt$ vertices, as either polarisation preferentially
selects one or the other of the $Z'_2$ and $Z'_3$ contributions in $A_L$, more so in the case of the 4DCHM interferences
with the SM that the 4DCHM contributions alone.

\begin{figure}[h!]
\begin{center}
\hspace{-0.5cm}
\includegraphics[width=0.45\textwidth]{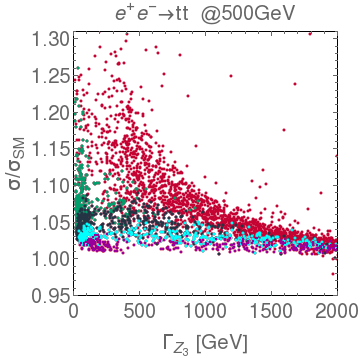}
\vspace{-0.1cm}
\caption{\small \it 
Predicted deviations for the cross section of the process $e^+e^-\to t\bar t$ at 500 GeV in the 4DCHM with respect to the SM as function of the $Z'_3$ width. The points correspond to $f=0.75-1.5$ TeV, $g_\rho=1.5-3$. The colour code is 
as follows. Red: all points; green: bounds on extra fermions as in Fig.~\ref{tLR}; black, cyan and purple are as green with
an additional bound on $M_{Z'}>$ 2, 2.5 and 3~TeV, respectively.}
\label{xs_GammaZ3_500}
\end{center}
\end{figure}

In order to further understand the dependence of the deviations induced by 4DCHM effects onto the SM upon the
contribution of $Z'$ mediation, 
 we plot in Fig.~\ref{xs_GammaZ3_500} the ratios of the cross section in the two models 
 at $\sqrt{s}=500$ GeV as function of the $Z'_3$ width. We use different colours for the contributions coming from different $Z'$ masses. It is clear that even for masses $M_{Z'}>5 \sqrt{s}$ we get $\sim$ 5\% deviations in the cross section coming from the interference while the square of the $Z'_3$ contribution is only sizable at small masses. Finally, the
effects are modulated somewhat by the $t^\prime$ mass, as this governs to some extent the actual size of the $Z'_3$ width.
(The plot for the $Z'_2$ case is rather similar, so we refrained from presenting it here.)

\section{\label{conclude} Conclusions}

To summarise, we have exploited a calculable version of a CHM, the so-called 4DCHM, which allows one to generate
the full spectrum of masses and couplings of all the NP particles present in it alongside relevant mixing with the 
SM states, in order to test the sensitivity of future $e^+e^-$ colliders to deviations in the cross section and (charge and spin) asymmetries of $t\bar t$ production from the SM values. Such a NP scenario, unlike more rudimentary implementations of CHMs, accounts for both the rescaling of the $Ztt$ coupling and the presence of the $Z'tt$ ones (specifically of three such states although only two play a phenomenologically significant role), which enter the physical
observables via interferences with the SM as well as like resonances (albeit through their tails, as we have assumed 
$\sqrt s<M_{Z'}$, whatever the $Z'$ state).

 Effects are sizable in several such observables, both at exclusive and inclusive level, 
 irrespectively of the CM energy of the collider, though they typically
grow with the latter owing to the presence of such additional neutral gauge bosons in the $s$-channel propagators, to
the extent that they are in general larger than the deviations induced by the rescaling of the $Ztt$ coupling. Therefore, altogether, such phenomena should be accessible at any realisation of a future $e^+e^-$ collider
currently considered, with or without polarisation of the beams. In fact, such a feature is not a key to their extraction,
rather it would serve the purpose of disentangling the various 4DCHM components in action. Contrast this with the fact
that the 4DCHM would in general (i.e., over most of its parameter space) escape the scope of the LHC, both at standard
and high luminosity, as neither the additional gauge bosons nor the additional fermions present in such a CHM may be accessible at the CERN machine, either directly (i.e., as visible objects) or indirectly (i.e., through effects onto SM observables). 

Further, under these circumstances, wherein a detection of new $Z'$s from a CHM cannot either be established with enough significance at the LHC or else the CERN machine cannot resolve nearby resonances (as typically predicted by theories like the one considered here, of pNGBs emerging from a $SO(5)\to SO(4)$ breaking), future leptonic colliders also afford one
with the ability to combine the aforementioned (charge and spin) 
 asymmetries together with the total cross section for the process $e^+e^-\to t \bar t$, thereby enabling one 
to increase significances, to the extent of possibly claiming an indirect discovery of a CHM structure of EWSB, even for
$Z'$ masses well beyond the kinematic reach of the leptonic accelerator. 

In short, realistic CHMs, wherein the additional spin-1 and spin-1/2 states are not integrated out as customarily done
in more simplistic realisations, are prime candidates for experimental scrutiny even beyond the LHC era, should this fail
to reveal such new objects. We have based our conclusions on a numerical study performed with on-shell top (anti)quarks, so they should eventually be validated by a proper analysis which accounts for their decay as well as parton shower and hadronisation in presence of detector effects. However, we are confident that, thanks to the cleanliness of 
a leptonic collider and its consequent efficiency in reconstructing the (anti)top quarks, the most salient features of our results will be preserved. To eventually enable such more sophisticated studies, 
a full implementation of the 4DCHM is available on the HEPMDB.
 
\section*{Acknowledgements}
SM is financially supported in part by the NExT Institute and further acknowledges funding from the Japan Society for the Promotion of Science (JSPS) while part of this research was carried out in the form of a Short Term Fellowship for Research in Japan (Grant Number S14026). He is also grateful to the Theoretical Physics Group at the Department of Physics of the University of Toyama for their kind hospitality during the tenure of the JSPS award. The work of GMP has been supported by the Swiss National Science Foundation (SNF) under contract 200021-160156. The authors are grateful to Patrick Janot and Francois Richard for helpful comments and suggestions, and to the hospitality of the Galileo Galilei Institute in Florence, where part of this work has been carried out during the workshop ``Prospects and Precision at the LHC at 14 TeV''.
The coordinates for the FCC-ee contour appeared in Figure~\ref{grojean} were provided by Patrick Janot via private communication, to which the authors are especially grateful.

\newpage

\bibliographystyle{JHEP}
\bibliography{4DCHMLC}

\end{document}